\ifdefined\pdfoutput\pdfoutput=1\fi
\PassOptionsToPackage{dvipsnames}{xcolor}

\documentclass[a4paper,onecolumn,11pt,unpublished,nopdfoutputerror]{quantumarticle}
\usepackage[T1]{fontenc}
\usepackage{amsmath,amssymb,booktabs,tabularx,array}
\usepackage{xcolor}
\usepackage[numbers,sort&compress]{natbib}
\usepackage{hyperref,xurl}
\providecommand{\doi}[1]{\href{https://doi.org/#1}{doi:\nolinkurl{#1}}}
\hypersetup{colorlinks=true,linkcolor=MidnightBlue,citecolor=MidnightBlue,urlcolor=MidnightBlue}
\usepackage{graphicx,tikz,pgfplots,placeins}
\usetikzlibrary{arrows.meta,positioning,calc}
\pgfplotsset{compat=1.18}
\definecolor{Ink}{HTML}{18334D}
\definecolor{Sea}{HTML}{087F8C}
\definecolor{Gold}{HTML}{BA641E}
\definecolor{Pale}{HTML}{EEF5F7}
\tikzset{stage/.style={draw=Ink!45,rounded corners=3pt,fill=Pale,
 text width=4.1cm,minimum height=1.25cm,align=flush center,font=\footnotesize,inner sep=5pt},
 flow/.style={-{Stealth[length=2.2mm]},thick,draw=Ink},
 note/.style={font=\footnotesize,align=center,text=Ink}}
\pgfplotsset{paperaxis/.style={font=\small,axis line style={Ink!40},
 tick style={Ink!40},tick label style={font=\footnotesize},
 grid style={Ink!10},axis on top,scaled ticks=false}}
\newcommand{\TQ}{\ensuremath{N_{\mathrm{Tof}}Q_{\max}}}
\newtheorem{theorem}{Theorem}
\newtheorem{corollary}[theorem]{Corollary}

\newcommand{\GenJoint}{\ensuremath{4.94\times10^{-6}}}

\newcommand{\ExactFull}{\ensuremath{6.248\times10^{-8}}}

\newcommand{\ExactLost}{\ensuremath{2.6\times10^{-14}}}

\newcommand{\ExactBias}{\ensuremath{-9.566\times10^{-8}}}
\newcommand{\ExactBiasPrice}{\ensuremath{-2.625\times10^{-6}}}
\newcommand{\TieN}{\ensuremath{2\times10^{9}}}
\newcommand{\TieSensN}{\ensuremath{2\times10^{8}}}
\newcommand{\TieNear}{35,508}
\newcommand{\TieCons}{2,086}
\newcommand{\TieConsExit}{1,992}
\newcommand{\TieConsNoExit}{94}
\newcommand{\TieExitMinDtheta}{\ensuremath{1.65\times10^{-3}}}
\newcommand{\TieNoExitMaxDtheta}{\ensuremath{2.1\times10^{-5}}}
\newcommand{\TieWin}{\ensuremath{1.08\times10^{-6}}}
\newcommand{\TieWinHalf}{\ensuremath{4.87\times10^{-7}}}
\newcommand{\TieRef}{\ensuremath{2.37\times10^{-9}}}
\newcommand{\TieSens}{\ensuremath{2.61\times10^{-5}}}
\newcommand{\TieWinSum}{\ensuremath{1.6\times10^{-6}}}
\newcommand{\TieRefSum}{\ensuremath{7.6\times10^{-9}}}
\newcommand{\TarfTruthUpper}{\ensuremath{2.99573\times10^{-6}}}
\newcommand{\TarfWindowUpper}{\ensuremath{4.951045\times10^{-7}}}

\usepackage{xltabular,colortbl,enumitem,needspace}
\usepackage[most]{tcolorbox}
\usetikzlibrary{fit,backgrounds,decorations.pathreplacing}
\usepgfplotslibrary{groupplots}

\newtcolorbox{cardbox}{enhanced,colback=Pale,colframe=Ink!30,boxrule=0.4pt,arc=2pt,
 left=6pt,right=6pt,top=3pt,bottom=3pt,fontupper=\footnotesize,before skip=4pt,after skip=6pt}

\pgfplotsset{suppaxis/.style={font=\small,axis line style={Ink!40},tick style={Ink!40},tick label style={font=\footnotesize},grid style={Ink!10},axis on top,scaled ticks=false}}
\tikzset{tag/.style={draw=Sea,fill=white,text=Sea,rounded corners=5pt,font=\scriptsize\bfseries,inner sep=1.6pt},cost/.style={font=\scriptsize,text=Ink!80,align=center},reg/.style={draw=Ink!70,fill=Sea!25,minimum height=0.55cm,inner sep=0pt,font=\scriptsize},regalt/.style={draw=Ink!70,fill=Gold!35,minimum height=0.55cm,inner sep=0pt,font=\scriptsize}}

\makeatletter
\newcommand{\supp@get}[1]{\ifcsname supp@#1\endcsname\csname supp@#1\endcsname\else\PackageError{supplement}{No generated value #1}{Run scripts/build_supplement.py}\fi}
\newcommand{\rec}[2]{\supp@get{rec:#1:#2}}
\newcommand{\prof}[2]{\supp@get{prof:#1:#2}}
\newcommand{\peak}[2]{\supp@get{peak:#1:#2}}
\newcommand{\sub}[1]{\supp@get{sub:#1}}
\newcommand{\dv}[1]{\supp@get{dv:#1}}
\newcommand{\sq}[1]{``\supp@get{sq:#1}\kern.07em''}
\makeatother
\expandafter\def\csname supp@rec:autocall/F1-001:T\endcsname{8,521}
\expandafter\def\csname supp@rec:autocall/F1-001:Q\endcsname{482}
\expandafter\def\csname supp@rec:autocall/F1-001:TQ\endcsname{4,107,122}
\expandafter\def\csname supp@rec:autocall/F2-001:T\endcsname{9,487}
\expandafter\def\csname supp@rec:autocall/F2-001:Q\endcsname{477}
\expandafter\def\csname supp@rec:autocall/F2-001:TQ\endcsname{4,525,299}
\expandafter\def\csname supp@rec:autocall/F3-001:T\endcsname{7,551}
\expandafter\def\csname supp@rec:autocall/F3-001:Q\endcsname{518}
\expandafter\def\csname supp@rec:autocall/F3-001:TQ\endcsname{3,911,418}
\expandafter\def\csname supp@rec:autocall/F3-001s:T\endcsname{7,551}
\expandafter\def\csname supp@rec:autocall/F3-001s:Q\endcsname{518}
\expandafter\def\csname supp@rec:autocall/F3-001s:TQ\endcsname{3,911,418}
\expandafter\def\csname supp@rec:autocall/F4-001:T\endcsname{8,641}
\expandafter\def\csname supp@rec:autocall/F4-001:Q\endcsname{474}
\expandafter\def\csname supp@rec:autocall/F4-001:TQ\endcsname{4,095,834}
\expandafter\def\csname supp@rec:autocall/STD-004:T\endcsname{19,074}
\expandafter\def\csname supp@rec:autocall/STD-004:Q\endcsname{610}
\expandafter\def\csname supp@rec:autocall/STD-004:TQ\endcsname{11,635,140}
\expandafter\def\csname supp@rec:autocall/v1:T\endcsname{16,810}
\expandafter\def\csname supp@rec:autocall/v1:Q\endcsname{556}
\expandafter\def\csname supp@rec:autocall/v1:TQ\endcsname{9,346,360}
\expandafter\def\csname supp@rec:hhl/F1-001:T\endcsname{1,482}
\expandafter\def\csname supp@rec:hhl/F1-001:Q\endcsname{151}
\expandafter\def\csname supp@rec:hhl/F1-001:TQ\endcsname{223,782}
\expandafter\def\csname supp@rec:hhl/F1-002:T\endcsname{1,057}
\expandafter\def\csname supp@rec:hhl/F1-002:Q\endcsname{126}
\expandafter\def\csname supp@rec:hhl/F1-002:TQ\endcsname{133,182}
\expandafter\def\csname supp@rec:hhl/F1-002s:T\endcsname{1,057}
\expandafter\def\csname supp@rec:hhl/F1-002s:Q\endcsname{126}
\expandafter\def\csname supp@rec:hhl/F1-002s:TQ\endcsname{133,182}
\expandafter\def\csname supp@rec:hhl/F2-001:T\endcsname{18,837}
\expandafter\def\csname supp@rec:hhl/F2-001:Q\endcsname{97}
\expandafter\def\csname supp@rec:hhl/F2-001:TQ\endcsname{1,827,189}
\expandafter\def\csname supp@rec:hhl/F3-001:T\endcsname{2,537}
\expandafter\def\csname supp@rec:hhl/F3-001:Q\endcsname{178}
\expandafter\def\csname supp@rec:hhl/F3-001:TQ\endcsname{451,586}
\expandafter\def\csname supp@rec:hhl/F4-001:T\endcsname{2,903}
\expandafter\def\csname supp@rec:hhl/F4-001:Q\endcsname{152}
\expandafter\def\csname supp@rec:hhl/F4-001:TQ\endcsname{441,256}
\expandafter\def\csname supp@rec:hhl/base:T\endcsname{14,922}
\expandafter\def\csname supp@rec:hhl/base:Q\endcsname{323}
\expandafter\def\csname supp@rec:hhl/base:TQ\endcsname{4,819,806}
\expandafter\def\csname supp@rec:hhl/gu:T\endcsname{1,348}
\expandafter\def\csname supp@rec:hhl/gu:Q\endcsname{115}
\expandafter\def\csname supp@rec:hhl/gu:TQ\endcsname{155,020}
\expandafter\def\csname supp@rec:hhl/v1:T\endcsname{2,163}
\expandafter\def\csname supp@rec:hhl/v1:Q\endcsname{193}
\expandafter\def\csname supp@rec:hhl/v1:TQ\endcsname{417,459}
\expandafter\def\csname supp@rec:m4a/F3G-001:T\endcsname{81,552}
\expandafter\def\csname supp@rec:m4a/F3G-001:Q\endcsname{640}
\expandafter\def\csname supp@rec:m4a/F3G-001:TQ\endcsname{52,193,280}
\expandafter\def\csname supp@rec:m4a/STDG-004:T\endcsname{18,822}
\expandafter\def\csname supp@rec:m4a/STDG-004:Q\endcsname{610}
\expandafter\def\csname supp@rec:m4a/STDG-004:TQ\endcsname{11,481,420}
\expandafter\def\csname supp@rec:m4a/V1G-001:T\endcsname{16,810}
\expandafter\def\csname supp@rec:m4a/V1G-001:Q\endcsname{556}
\expandafter\def\csname supp@rec:m4a/V1G-001:TQ\endcsname{9,346,360}
\expandafter\def\csname supp@rec:m4b/F1-001:T\endcsname{13,661}
\expandafter\def\csname supp@rec:m4b/F1-001:Q\endcsname{524}
\expandafter\def\csname supp@rec:m4b/F1-001:TQ\endcsname{7,158,364}
\expandafter\def\csname supp@rec:m4b/F1-001s:T\endcsname{13,661}
\expandafter\def\csname supp@rec:m4b/F1-001s:Q\endcsname{524}
\expandafter\def\csname supp@rec:m4b/F1-001s:TQ\endcsname{7,158,364}
\expandafter\def\csname supp@rec:m4b/F2-001:T\endcsname{12,685}
\expandafter\def\csname supp@rec:m4b/F2-001:Q\endcsname{547}
\expandafter\def\csname supp@rec:m4b/F2-001:TQ\endcsname{6,938,695}
\expandafter\def\csname supp@rec:m4b/F2-001s:T\endcsname{12,685}
\expandafter\def\csname supp@rec:m4b/F2-001s:Q\endcsname{547}
\expandafter\def\csname supp@rec:m4b/F2-001s:TQ\endcsname{6,938,695}
\expandafter\def\csname supp@rec:tarf/F1-001:T\endcsname{10,850}
\expandafter\def\csname supp@rec:tarf/F1-001:Q\endcsname{297}
\expandafter\def\csname supp@rec:tarf/F1-001:TQ\endcsname{3,222,450}
\expandafter\def\csname supp@rec:tarf/F2-001:T\endcsname{11,805}
\expandafter\def\csname supp@rec:tarf/F2-001:Q\endcsname{301}
\expandafter\def\csname supp@rec:tarf/F2-001:TQ\endcsname{3,553,305}
\expandafter\def\csname supp@rec:tarf/F3-001:T\endcsname{10,040}
\expandafter\def\csname supp@rec:tarf/F3-001:Q\endcsname{286}
\expandafter\def\csname supp@rec:tarf/F3-001:TQ\endcsname{2,871,440}
\expandafter\def\csname supp@rec:tarf/F3-002:T\endcsname{8,381}
\expandafter\def\csname supp@rec:tarf/F3-002:Q\endcsname{269}
\expandafter\def\csname supp@rec:tarf/F3-002:TQ\endcsname{2,254,489}
\expandafter\def\csname supp@rec:tarf/F3-002s:T\endcsname{8,381}
\expandafter\def\csname supp@rec:tarf/F3-002s:Q\endcsname{269}
\expandafter\def\csname supp@rec:tarf/F3-002s:TQ\endcsname{2,254,489}
\expandafter\def\csname supp@rec:tarf/F4-001:T\endcsname{9,874}
\expandafter\def\csname supp@rec:tarf/F4-001:Q\endcsname{280}
\expandafter\def\csname supp@rec:tarf/F4-001:TQ\endcsname{2,764,720}
\expandafter\def\csname supp@rec:tarf/F4-002:T\endcsname{8,894}
\expandafter\def\csname supp@rec:tarf/F4-002:Q\endcsname{270}
\expandafter\def\csname supp@rec:tarf/F4-002:TQ\endcsname{2,401,380}
\expandafter\def\csname supp@rec:tarf/price:T\endcsname{3,334,652}
\expandafter\def\csname supp@rec:tarf/price:Q\endcsname{653}
\expandafter\def\csname supp@rec:tarf/price:TQ\endcsname{2,177,527,756}
\expandafter\def\csname supp@rec:tarf/v1:T\endcsname{28,428}
\expandafter\def\csname supp@rec:tarf/v1:Q\endcsname{384}
\expandafter\def\csname supp@rec:tarf/v1:TQ\endcsname{10,916,352}
\expandafter\def\csname supp@prof:autocall/F1-001:plookup\endcsname{2,400}
\expandafter\def\csname supp@peak:autocall/F1-001:plookup\endcsname{461}
\expandafter\def\csname supp@prof:autocall/F1-001:pclean\endcsname{400}
\expandafter\def\csname supp@peak:autocall/F1-001:pclean\endcsname{424}
\expandafter\def\csname supp@prof:autocall/F1-001:padds\endcsname{4,178}
\expandafter\def\csname supp@peak:autocall/F1-001:padds\endcsname{481}
\expandafter\def\csname supp@prof:autocall/F1-001:tests\endcsname{700}
\expandafter\def\csname supp@peak:autocall/F1-001:tests\endcsname{437}
\expandafter\def\csname supp@prof:autocall/F1-001:pother\endcsname{843}
\expandafter\def\csname supp@peak:autocall/F1-001:pother\endcsname{482}
\expandafter\def\csname supp@prof:autocall/F3-001:padds\endcsname{894}
\expandafter\def\csname supp@peak:autocall/F3-001:padds\endcsname{376}
\expandafter\def\csname supp@prof:autocall/F3-001:forms\endcsname{4,398}
\expandafter\def\csname supp@peak:autocall/F3-001:forms\endcsname{496}
\expandafter\def\csname supp@prof:autocall/F3-001:tests\endcsname{884}
\expandafter\def\csname supp@peak:autocall/F3-001:tests\endcsname{518}
\expandafter\def\csname supp@prof:autocall/F3-001:pworst\endcsname{460}
\expandafter\def\csname supp@peak:autocall/F3-001:pworst\endcsname{500}
\expandafter\def\csname supp@prof:autocall/F3-001:pother\endcsname{915}
\expandafter\def\csname supp@peak:autocall/F3-001:pother\endcsname{497}
\expandafter\def\csname supp@prof:autocall/v1:plookup\endcsname{3,600}
\expandafter\def\csname supp@peak:autocall/v1:plookup\endcsname{531}
\expandafter\def\csname supp@prof:autocall/v1:pclean\endcsname{3,600}
\expandafter\def\csname supp@peak:autocall/v1:pclean\endcsname{531}
\expandafter\def\csname supp@prof:autocall/v1:padds\endcsname{6,960}
\expandafter\def\csname supp@peak:autocall/v1:padds\endcsname{556}
\expandafter\def\csname supp@prof:autocall/v1:tests\endcsname{1,000}
\expandafter\def\csname supp@peak:autocall/v1:tests\endcsname{555}
\expandafter\def\csname supp@prof:autocall/v1:pother\endcsname{1,650}
\expandafter\def\csname supp@peak:autocall/v1:pother\endcsname{547}
\expandafter\def\csname supp@prof:hhl/F1-001:lookup\endcsname{240}
\expandafter\def\csname supp@peak:hhl/F1-001:lookup\endcsname{130}
\expandafter\def\csname supp@prof:hhl/F1-001:unlookup\endcsname{240}
\expandafter\def\csname supp@peak:hhl/F1-001:unlookup\endcsname{124}
\expandafter\def\csname supp@prof:hhl/F1-001:norm\endcsname{83}
\expandafter\def\csname supp@peak:hhl/F1-001:norm\endcsname{--}
\expandafter\def\csname supp@prof:hhl/F1-001:cadd\endcsname{32}
\expandafter\def\csname supp@peak:hhl/F1-001:cadd\endcsname{151}
\expandafter\def\csname supp@prof:hhl/F1-001:crows\endcsname{336}
\expandafter\def\csname supp@peak:hhl/F1-001:crows\endcsname{127}
\expandafter\def\csname supp@prof:hhl/F1-001:mrows\endcsname{551}
\expandafter\def\csname supp@peak:hhl/F1-001:mrows\endcsname{145}
\expandafter\def\csname supp@prof:hhl/F1-002:lookup\endcsname{221}
\expandafter\def\csname supp@peak:hhl/F1-002:lookup\endcsname{120}
\expandafter\def\csname supp@prof:hhl/F1-002:unlookup\endcsname{124}
\expandafter\def\csname supp@peak:hhl/F1-002:unlookup\endcsname{92}
\expandafter\def\csname supp@prof:hhl/F1-002:norm\endcsname{76}
\expandafter\def\csname supp@peak:hhl/F1-002:norm\endcsname{--}
\expandafter\def\csname supp@prof:hhl/F1-002:cadd\endcsname{41}
\expandafter\def\csname supp@peak:hhl/F1-002:cadd\endcsname{126}
\expandafter\def\csname supp@prof:hhl/F1-002:crows\endcsname{226}
\expandafter\def\csname supp@peak:hhl/F1-002:crows\endcsname{105}
\expandafter\def\csname supp@prof:hhl/F1-002:mrows\endcsname{369}
\expandafter\def\csname supp@peak:hhl/F1-002:mrows\endcsname{126}
\expandafter\def\csname supp@prof:tarf/F1-001:index\endcsname{400}
\expandafter\def\csname supp@peak:tarf/F1-001:index\endcsname{229}
\expandafter\def\csname supp@prof:tarf/F1-001:lookup\endcsname{3,926}
\expandafter\def\csname supp@peak:tarf/F1-001:lookup\endcsname{278}
\expandafter\def\csname supp@prof:tarf/F1-001:clean\endcsname{1,032}
\expandafter\def\csname supp@peak:tarf/F1-001:clean\endcsname{235}
\expandafter\def\csname supp@prof:tarf/F1-001:adds\endcsname{2,704}
\expandafter\def\csname supp@peak:tarf/F1-001:adds\endcsname{297}
\expandafter\def\csname supp@prof:tarf/F1-001:stop\endcsname{232}
\expandafter\def\csname supp@peak:tarf/F1-001:stop\endcsname{225}
\expandafter\def\csname supp@prof:tarf/F1-001:cap\endcsname{1,214}
\expandafter\def\csname supp@peak:tarf/F1-001:cap\endcsname{278}
\expandafter\def\csname supp@prof:tarf/F1-001:sqrt\endcsname{456}
\expandafter\def\csname supp@peak:tarf/F1-001:sqrt\endcsname{272}
\expandafter\def\csname supp@prof:tarf/F1-001:angle\endcsname{886}
\expandafter\def\csname supp@peak:tarf/F1-001:angle\endcsname{290}
\expandafter\def\csname supp@prof:tarf/F3-001:index\endcsname{400}
\expandafter\def\csname supp@peak:tarf/F3-001:index\endcsname{227}
\expandafter\def\csname supp@prof:tarf/F3-001:lookup\endcsname{3,898}
\expandafter\def\csname supp@peak:tarf/F3-001:lookup\endcsname{274}
\expandafter\def\csname supp@prof:tarf/F3-001:clean\endcsname{510}
\expandafter\def\csname supp@peak:tarf/F3-001:clean\endcsname{232}
\expandafter\def\csname supp@prof:tarf/F3-001:adds\endcsname{2,500}
\expandafter\def\csname supp@peak:tarf/F3-001:adds\endcsname{286}
\expandafter\def\csname supp@prof:tarf/F3-001:stop\endcsname{148}
\expandafter\def\csname supp@peak:tarf/F3-001:stop\endcsname{223}
\expandafter\def\csname supp@prof:tarf/F3-001:cap\endcsname{1,289}
\expandafter\def\csname supp@peak:tarf/F3-001:cap\endcsname{283}
\expandafter\def\csname supp@prof:tarf/F3-001:sqrt\endcsname{414}
\expandafter\def\csname supp@peak:tarf/F3-001:sqrt\endcsname{271}
\expandafter\def\csname supp@prof:tarf/F3-001:angle\endcsname{881}
\expandafter\def\csname supp@peak:tarf/F3-001:angle\endcsname{285}
\expandafter\def\csname supp@prof:tarf/F3-002:index\endcsname{300}
\expandafter\def\csname supp@peak:tarf/F3-002:index\endcsname{218}
\expandafter\def\csname supp@prof:tarf/F3-002:lookup\endcsname{3,738}
\expandafter\def\csname supp@peak:tarf/F3-002:lookup\endcsname{259}
\expandafter\def\csname supp@prof:tarf/F3-002:clean\endcsname{453}
\expandafter\def\csname supp@peak:tarf/F3-002:clean\endcsname{224}
\expandafter\def\csname supp@prof:tarf/F3-002:adds\endcsname{2,072}
\expandafter\def\csname supp@peak:tarf/F3-002:adds\endcsname{269}
\expandafter\def\csname supp@prof:tarf/F3-002:stop\endcsname{85}
\expandafter\def\csname supp@peak:tarf/F3-002:stop\endcsname{216}
\expandafter\def\csname supp@prof:tarf/F3-002:cap\endcsname{778}
\expandafter\def\csname supp@peak:tarf/F3-002:cap\endcsname{266}
\expandafter\def\csname supp@prof:tarf/F3-002:sqrt\endcsname{414}
\expandafter\def\csname supp@peak:tarf/F3-002:sqrt\endcsname{261}
\expandafter\def\csname supp@prof:tarf/F3-002:angle\endcsname{541}
\expandafter\def\csname supp@peak:tarf/F3-002:angle\endcsname{269}
\expandafter\def\csname supp@prof:tarf/F4-001:index\endcsname{400}
\expandafter\def\csname supp@peak:tarf/F4-001:index\endcsname{224}
\expandafter\def\csname supp@prof:tarf/F4-001:lookup\endcsname{3,898}
\expandafter\def\csname supp@peak:tarf/F4-001:lookup\endcsname{268}
\expandafter\def\csname supp@prof:tarf/F4-001:clean\endcsname{510}
\expandafter\def\csname supp@peak:tarf/F4-001:clean\endcsname{229}
\expandafter\def\csname supp@prof:tarf/F4-001:adds\endcsname{2,350}
\expandafter\def\csname supp@peak:tarf/F4-001:adds\endcsname{280}
\expandafter\def\csname supp@prof:tarf/F4-001:stop\endcsname{78}
\expandafter\def\csname supp@peak:tarf/F4-001:stop\endcsname{218}
\expandafter\def\csname supp@prof:tarf/F4-001:cap\endcsname{1,277}
\expandafter\def\csname supp@peak:tarf/F4-001:cap\endcsname{280}
\expandafter\def\csname supp@prof:tarf/F4-001:sqrt\endcsname{414}
\expandafter\def\csname supp@peak:tarf/F4-001:sqrt\endcsname{271}
\expandafter\def\csname supp@prof:tarf/F4-001:angle\endcsname{947}
\expandafter\def\csname supp@peak:tarf/F4-001:angle\endcsname{280}
\expandafter\def\csname supp@prof:tarf/F4-002:index\endcsname{300}
\expandafter\def\csname supp@peak:tarf/F4-002:index\endcsname{219}
\expandafter\def\csname supp@prof:tarf/F4-002:lookup\endcsname{3,738}
\expandafter\def\csname supp@peak:tarf/F4-002:lookup\endcsname{259}
\expandafter\def\csname supp@prof:tarf/F4-002:clean\endcsname{453}
\expandafter\def\csname supp@peak:tarf/F4-002:clean\endcsname{224}
\expandafter\def\csname supp@prof:tarf/F4-002:adds\endcsname{2,072}
\expandafter\def\csname supp@peak:tarf/F4-002:adds\endcsname{270}
\expandafter\def\csname supp@prof:tarf/F4-002:stop\endcsname{85}
\expandafter\def\csname supp@peak:tarf/F4-002:stop\endcsname{216}
\expandafter\def\csname supp@prof:tarf/F4-002:cap\endcsname{1,176}
\expandafter\def\csname supp@peak:tarf/F4-002:cap\endcsname{270}
\expandafter\def\csname supp@prof:tarf/F4-002:sqrt\endcsname{414}
\expandafter\def\csname supp@peak:tarf/F4-002:sqrt\endcsname{261}
\expandafter\def\csname supp@prof:tarf/F4-002:angle\endcsname{656}
\expandafter\def\csname supp@peak:tarf/F4-002:angle\endcsname{270}
\expandafter\def\csname supp@prof:tarf/v1:index\endcsname{676}
\expandafter\def\csname supp@peak:tarf/v1:index\endcsname{317}
\expandafter\def\csname supp@prof:tarf/v1:lookup\endcsname{4,562}
\expandafter\def\csname supp@peak:tarf/v1:lookup\endcsname{321}
\expandafter\def\csname supp@prof:tarf/v1:clean\endcsname{4,562}
\expandafter\def\csname supp@peak:tarf/v1:clean\endcsname{321}
\expandafter\def\csname supp@prof:tarf/v1:adds\endcsname{7,384}
\expandafter\def\csname supp@peak:tarf/v1:adds\endcsname{384}
\expandafter\def\csname supp@prof:tarf/v1:stop\endcsname{5,352}
\expandafter\def\csname supp@peak:tarf/v1:stop\endcsname{351}
\expandafter\def\csname supp@prof:tarf/v1:cap\endcsname{2,058}
\expandafter\def\csname supp@peak:tarf/v1:cap\endcsname{322}
\expandafter\def\csname supp@prof:tarf/v1:angle\endcsname{3,834}
\expandafter\def\csname supp@peak:tarf/v1:angle\endcsname{351}
\expandafter\def\csname supp@sub:autoFthreeRone\endcsname{1,764}
\expandafter\def\csname supp@sub:autoFthreeRtwo\endcsname{2,634}
\expandafter\def\csname supp@sub:autoFthreeCallForms\endcsname{430}
\expandafter\def\csname supp@sub:autoFthreeUadds\endcsname{894}
\expandafter\def\csname supp@sub:tarfFoneCleanFwd\endcsname{516}
\expandafter\def\csname supp@sub:tarfFthreeCleanFwd\endcsname{0}
\expandafter\def\csname supp@sub:tarfFthreeStopRev\endcsname{100}
\expandafter\def\csname supp@sub:tarfFfourStopRev\endcsname{30}
\expandafter\def\csname supp@sub:tarfFfourTwoCapMul\endcsname{1,049}
\expandafter\def\csname supp@sub:tarfFthreeTwoCapMul\endcsname{651}
\expandafter\def\csname supp@sub:tarfVoneCompare\endcsname{4,544}
\expandafter\def\csname supp@dv:autoDatesVone\endcsname{15,160}
\expandafter\def\csname supp@dv:autoDatesFone\endcsname{7,678}
\expandafter\def\csname supp@dv:autoDatesFthree\endcsname{6,176}
\expandafter\def\csname supp@dv:autoPathVone\endcsname{14,160}
\expandafter\def\csname supp@dv:autoPathFone\endcsname{6,978}
\expandafter\def\csname supp@dv:autoPathShareVone\endcsname{84}
\expandafter\def\csname supp@dv:autoFormsPlusAdds\endcsname{5,292}
\expandafter\def\csname supp@dv:autoPayFthree\endcsname{1,375}
\expandafter\def\csname supp@dv:autoTQvoneFthree\endcsname{2.39}
\expandafter\def\csname supp@dv:autoTQfoneFthree\endcsname{5}
\expandafter\def\csname supp@dv:autoDatesFoneFthree\endcsname{20}
\expandafter\def\csname supp@dv:hhlTQgain\endcsname{40}
\expandafter\def\csname supp@dv:hhlTgain\endcsname{425}
\expandafter\def\csname supp@dv:hhlQgain\endcsname{25}
\expandafter\def\csname supp@dv:hhlVsGu\endcsname{14.1}
\expandafter\def\csname supp@dv:hhlTableRatio\endcsname{13.7}
\expandafter\def\csname supp@dv:tarfDatesVone\endcsname{22,536}
\expandafter\def\csname supp@dv:tarfDatesFone\endcsname{8,294}
\expandafter\def\csname supp@dv:tarfDatesFthreeTwo\endcsname{6,648}
\expandafter\def\csname supp@dv:tarfPayVone\endcsname{5,892}
\expandafter\def\csname supp@dv:tarfPayFone\endcsname{2,556}
\expandafter\def\csname supp@dv:tarfPayFthreeTwo\endcsname{1,733}
\expandafter\def\csname supp@dv:tarfTQvoneFone\endcsname{3.39}
\expandafter\def\csname supp@dv:tarfTQfoneFthreeTwo\endcsname{1.43}
\expandafter\def\csname supp@dv:tarfTQvoneFthreeTwo\endcsname{4.84}
\expandafter\def\csname supp@dv:tarfTQpriceVone\endcsname{199.5}
\expandafter\def\csname supp@dv:tarfAngleSqrtFone\endcsname{1,342}
\expandafter\def\csname supp@dv:transferLoss\endcsname{5.58}
\expandafter\def\csname supp@dv:exactFull\endcsname{\ensuremath{6.248\times10^{-8}}}
\expandafter\def\csname supp@dv:genSA\endcsname{1.35}
\expandafter\def\csname supp@dv:genLU\endcsname{1.31}
\expandafter\def\csname supp@dv:genSAvsTransfer\endcsname{7.5}
\expandafter\def\csname supp@dv:genTailSA\endcsname{49}
\expandafter\def\csname supp@dv:genTailLU\endcsname{46}
\expandafter\def\csname supp@dv:hhlErrUnits\endcsname{417.45}
\expandafter\def\csname supp@dv:hhlErrRad\endcsname{\ensuremath{9.72\times10^{-8}}}
\expandafter\def\csname supp@dv:hhlTolUnits\endcsname{429.50}
\expandafter\def\csname supp@dv:hhlMargin\endcsname{12.0}
\expandafter\def\csname supp@dv:hhlWorstX\endcsname{151,614}
\expandafter\def\csname supp@dv:hhlSeconds\endcsname{29}
\expandafter\def\csname supp@sq:ftwoPack\endcsname{Packing beats sparsity}
\expandafter\def\csname supp@sq:hBooth\endcsname{Radix-4 Booth rows beat binary rows for products added into a wide target}
\expandafter\def\csname supp@sq:tDefer\endcsname{Forward-pass MBU fix-ups are free under Bennett reversal}
\expandafter\def\csname supp@sq:tReach\endcsname{Register widths are set by reachable ranges, not by the worst case}
\expandafter\def\csname supp@sq:tTop\endcsname{In a Gidney adder the carry into the top bit never needs an ancilla}
\expandafter\def\csname supp@sq:gTie\endcsname{when scoring a form, tau may only sit between DISTINCT form values}

\begin{document}
\title{Reducing the arithmetic cost of quantum derivative pricing with language-model circuit design}
\author{Monit Sharma}
\author{Hoong Chuin Lau}
\affiliation{Singapore Management University, Singapore}
\maketitle

\begin{abstract}
Quantum amplitude estimation can price derivatives with quadratically fewer oracle calls than classical Monte Carlo, but every call must evaluate the payoff coherently. We construct reversible payoff arithmetic for the three-asset worst-of autocallable and the target accrual redemption forward (TARF) benchmarks of \citet{chakrabarti}, using an evaluator-guided, AI-assisted design process. The circuits keep compact integer representations of market paths alongside contract history and reconstruct thresholds and payoff angles only temporarily. In the product of Toffoli count and peak logical qubits, they are 2.39- to 4.17-fold cheaper for the autocallable and 4.84- to 5.52-fold cheaper for TARF than references that already avoid coherent exponentials through log-return tests~\citep{wang} or payment lookups, and 1,250 to 2,180 times and 966 to 1,100 times cheaper than our conservative reconstructions of the price-space arithmetic of \citet{chakrabarti} under the common digital-angle interface and target tolerance. For two autocallable circuits we prove exact coherent execution of the integer specification for every input, accumulator and measurement record, and lattice calculations give numerical angle-tolerance-violation probabilities of \ExactFull{} and $3.56\times10^{-8}$ and the resulting pricing bias. The lowest-product autocallable is more sensitive to threshold inputs, whereas targeted hardening improves selected boundary cases while preserving resource savings; at a second parameter setting only setting-specific designs keep the saving. The counts cover payoff arithmetic only, excluding loading, payoff rotation and estimation.
\end{abstract}

\section{Introduction}
\label{sec:introduction}
Derivative pricing estimates an expected discounted payoff under a market model. Quantum amplitude estimation needs quadratically fewer oracle calls than classical Monte Carlo~\citep{montanaro,rebentrost,stamatopoulos2020}, but each call must prepare the path distribution and evaluate the payoff coherently. For path-dependent contracts this payoff arithmetic is a large part of the logical cost. In the resource analysis of \citet{chakrabarti}, every simulated log-return is mapped to a price by a coherent exponential before the contract tests and the payoff are evaluated. Other encodings reduce or avoid this arithmetic: log-return comparisons avoid exponentials at monitoring dates~\citep{wang}, quantum signal processing and integration-based loading encode nonlinear payoffs into amplitudes~\citep{qsp,integration}, and Fourier-series and arithmetic-free methods avoid explicit payoff circuits for suitable integrands~\citep{herbert,mcardle}.

We keep the digital payoff-angle interface of \citet{chakrabarti} and ask how far its cost can be reduced while both the coherent action and the financial approximation remain assessable. These are separate obligations: a circuit can implement its integer specification exactly and still disagree with the real contract near a threshold. The constructions use a simple property of the benchmarks. With constant coefficients, uniform dates and an equally spaced factor grid, every log-return is an affine function of cumulative index sums. A circuit can therefore retain a few narrow integer registers and the contract history, and reconstruct weighted forms, thresholds and the payoff angle only when they are needed.

The circuits were developed in an evaluator-guided design loop: AI agents proposed and implemented candidate programs, and a separate scorer validated them and counted their resources (Section~\ref{sec:process}). We report the resulting circuits and their verification; we make no comparative claim about the design process. Our contributions are:
\begin{enumerate}
\item (Constructions) Compact payoff arithmetic for both contracts: index-sum and barrier-centred integer forms for the autocallable, and a windowed index with an exact cap identity for TARF (Section~\ref{sec:constructions}).
\item (Resources) Relative to log-domain and payment-lookup references, which already avoid coherent exponentials, the product of Toffoli count and peak qubits falls 2.39- to 4.17-fold for the autocallable and 4.84- to 5.52-fold for TARF; relative to our reconstructions of the price-space arithmetic of \citet{chakrabarti} under the common digital-angle interface, it falls 1,250- to 2,180-fold and 966- to 1,100-fold (Section~\ref{sec:resources}).
\item (Verification) For two autocallable circuits, all-input proofs of the coherent integer action, a logical inverse, and numerical lattice calculations of the violation probability and the pricing bias (Section~\ref{sec:verification}). On the HHL angle task of \citet{gu}, a signed-Booth circuit with a proved logical action has a product 28\% below the best published circuit, with two fewer qubits (Appendix~\ref{app:hhl}).
\item (Limits) A tradeoff between cost and robustness near contract thresholds, and parameter dependence: a fixed transfer to a second setting loses its advantage, while setting-specific designs recover a 4.29-fold saving (Sections~\ref{sec:parameter-transfer} and~\ref{sec:robustness}).
\end{enumerate}
The comparisons cover payoff arithmetic after loading and before payoff rotation, under matched tasks, angle formats and target tolerances. Complete pricing implementations are outside the claims.

\paragraph{Related work.} The constructions combine measurement-uncomputed adders~\citep{gidney}, unary-iteration lookups~\citep{babbush2018,berry2019}, piecewise-polynomial arithmetic~\citep{hrs} and Booth recoding~\citep{booth,macsorley}. Published resource estimates for these contracts~\citep{chakrabarti,qsharp,qsp,integration} count complete pricing algorithms under different accuracy boundaries, so we compare against our reconstructions under the same angle format and target tolerance rather than against their totals. The design loop follows the evaluator-guided program evolution of \citet{gu}, related to AlphaEvolve~\citep{alphaevolve}; other machine-assisted approaches optimize tensor decompositions, local rewrites and synthesis heuristics~\citep{alphatensor,lizziobosco,sim,jang}. Kickmix circuits admit classical checks of basis values, phases and measurement-based cleanup~\citep{babbush2026,khattar2026}.

\section{Problem definition and resource model}
\label{sec:task}
Both benchmarks follow \citet{chakrabarti}. The autocallable can terminate on scheduled call dates; if it survives, a barrier breach activates a terminal loss linked to the worst-performing asset. The TARF accumulates signed payments and terminates at a knock-out barrier or an accrual target. Both contracts therefore need contract history as well as the current market state. Table~\ref{tab:contracts} lists the parameters.

\begin{table}[htbp]
\centering
\small
\caption{Benchmark parameters. All settings mature at one year and use five bits per factor index. Between the autocallable settings, only volatility and correlation change.}
\label{tab:contracts}
\setlength{\tabcolsep}{4pt}
\renewcommand{\arraystretch}{1.12}
\begin{tabularx}{\linewidth}{@{}l>{\centering\arraybackslash}X>{\centering\arraybackslash}X>{\centering\arraybackslash}p{1.15cm}@{}}
\toprule
Parameter & Autocallable: original & Autocallable: second & TARF\\
\midrule
Factors $d$ & 3 & 3 & 1\\
Dates $T$ & 20 & 20 & 26\\
Time step $\Delta t$ & $1/20$ & $1/20$ & $1/26$\\
Rate $r$ & 0.02 & 0.02 & 0.02\\
Volatility $\sigma$ & $(0.10,0.25,0.40)$ & $(0.13,0.22,0.37)$ & 0.40\\
Pairwise correlation $\rho$ & 0.50 & 0.35 & ---\\
Input qubits & 300 & 300 & 130\\
\bottomrule
\end{tabularx}
\end{table}

\paragraph{Loading law.} Each independent five-bit index $j\in\{0,\ldots,31\}$ selects a grid midpoint $z_j$ with weight $p_j$:
\begin{equation}
z_j=-5+(j+\tfrac12)\frac{10}{32},\qquad
p_j=\frac{\exp(-z_j^2/2)}{\sum_{\ell=0}^{31}\exp(-z_\ell^2/2)}.
\label{eq:loadlaw}
\end{equation}
These are normalized midpoint density weights, not Gaussian bin probabilities. For $T$ dates and $d$ factors, write $j_t^{(k)}$ for the index at date $t$ and factor $k$, and $\mathbf z=(j_t^{(k)})$ for the encoded path. The loading law and the computational target are
\begin{equation}
p_{\mathrm{load}}(\mathbf z)
=\prod_{t=1}^{T}\prod_{k=0}^{d-1}p_{j_t^{(k)}},\qquad
V_{\mathrm{grid}}=\sum_{\mathbf z\in\{0,\ldots,31\}^{Td}}p_{\mathrm{load}}(\mathbf z)f_{\mathrm{disc}}(\mathbf z),
\label{eq:grid-value}
\end{equation}
where $f_{\mathrm{disc}}$ is the discounted payoff defined by the contract rules below.

\paragraph{Autocallable.} For assets $i,k\in\{0,1,2\}$, let $L$ be the Cholesky factor of the covariance matrix $\Sigma_{ik}=\Delta t\,\rho_{ik}\sigma_i\sigma_k$, with $\rho_{ii}=1$ and $\rho_{ik}=\rho$ for $i\ne k$. The cumulative log-returns and their minimum are
\begin{equation}
C_i(t)=t(r-\sigma_i^2/2)\Delta t
+\sum_{s=1}^{t}\sum_{k=0}^{2}L_{ik}z_{j_s^{(k)}},
\qquad W_t=\min_i C_i(t).
\label{eq:autopath}
\end{equation}
At dates $t=4a$, $a=1,\ldots,5$, the first $a_*$ with $W_{4a_*}\ge\ln1.1$ calls the contract. Monitoring at every date sets a knock-in flag when $W_t<\ln0.7$. Calls take priority over the terminal put:
\begin{equation}
f_{\mathrm{disc}}=
\begin{cases}
2a_*e^{-r a_*/5},&\text{if a call occurs},\\
18(e^{W_{20}}-1)e^{-r},
&\text{if no call, knock-in, and }W_{20}<0,\\
0,&\text{otherwise}.
\end{cases}
\label{eq:autopayoff}
\end{equation}

\paragraph{TARF.} For this single-factor task write $j_s=j_s^{(0)}$. The price starts at $S_0=20$ and evolves as
\begin{equation}
S_t=20\exp\!\left[(r-\sigma^2/2)\frac{t}{26}+\frac{\sigma}{\sqrt{26}}\sum_{s=1}^{t}z_{j_s}\right].
\label{eq:tarfpath}
\end{equation}
The accrual starts at $P_0=0$. At each date $t=1,\ldots,26$, let $P_t$ be the accrual immediately before the date-$t$ payment; the contract first tests knock-out: $S_t\ge30$ terminates without payment. Otherwise the uncapped payment is $h(S_t)=S_t-20$ for $20\le S_t<30$, zero for $15\le S_t<20$, and $2(S_t-20)$ for $S_t<15$. If $P_t+h(S_t)\ge5$, the contract pays $q_t=5-P_t$ and terminates; otherwise it pays $q_t=h(S_t)$ and updates $P_{t+1}=P_t+q_t$. Negative payments reduce the accrual. With $d_t=e^{-rt/26}$, $f_{\mathrm{disc}}=\sum_t d_tq_t$ for the payments $q_t$. We use the twenty-six-date resource benchmark of \citet{chakrabarti}.

\subsection{Payoff-angle interface and accuracy}
For each contract, fixed constants $(H,D)$ normalize the payoff, $\widetilde f=(f_{\mathrm{disc}}+H)/D\in[0,1]$ (Appendix~\ref{app:normalization}). The digital-angle interface of \citet{chakrabarti} computes
\begin{equation}
a(\mathbf z)=\arcsin\sqrt{\widetilde f(\mathbf z)},\qquad
|\mathbf z\rangle|\theta\rangle|0\rangle
\longmapsto
|\mathbf z\rangle|\theta+\widehat A(\mathbf z)\bmod 2^{22}\rangle|0\rangle,
\label{eq:oracle}
\end{equation}
where $\widehat A(\mathbf z)$ is the circuit's integer increment and $\widehat a=2^{-20}\widehat A$ its angle. The incoming accumulator $\theta$ is arbitrary, the path is preserved, workspace starts and ends at zero, and the action must hold coherently independently of measurement outcomes. The pricing construction initializes $\theta=0$ before applying the rotation, whereas Eq.~\eqref{eq:oracle} states the circuit action for an arbitrary incoming $\theta$. An ideal rotation $R_y(2\widehat a)$ on a payoff flag then gives the effective value
\begin{equation}
\widehat V_{\mathrm{grid}}=D\,\mathbb E_{p_{\mathrm{load}}}[\sin^2\widehat a]-H,
\label{eq:effective-value}
\end{equation}
which amplitude estimation can estimate~\citep{stamatopoulos2020,chakrabarti}. The accuracy requirement is
\begin{equation}
\Pr_{\mathbf z\sim p_{\mathrm{load}}}\!\left[|\widehat a(\mathbf z)-a(\mathbf z)|>\delta_{\mathrm{ang}}\right]\le\varepsilon_{\mathrm{fail}},\qquad
\delta_{\mathrm{ang}}=\varepsilon_{\mathrm{fail}}=10^{-5}.
\label{eq:task}
\end{equation}
We call the event in Eq.~\eqref{eq:task} an \emph{angle-tolerance violation}. Input modification, dirty workspace and phase errors have no allowance.

\subsection{Resource objective}
For each contract, loading law and angle encoding, the objective is to reduce $\TQ$ subject to Eqs.~\eqref{eq:oracle} and~\eqref{eq:task}. Here $N_{\mathrm{Tof}}$ counts CCX, CCZ and controlled-SWAP operations in Toffoli-equivalent units, and $Q_{\max}$ counts peak live logical qubits, including inputs, output and workspace, with reuse after clean freeing. The component covers path arithmetic, contract decisions and the digital angle after loading and before payoff rotation. The counts are not compiled Clifford+$T$, depth or physical-qubit estimates.

\section{Circuit constructions}
\label{sec:constructions}
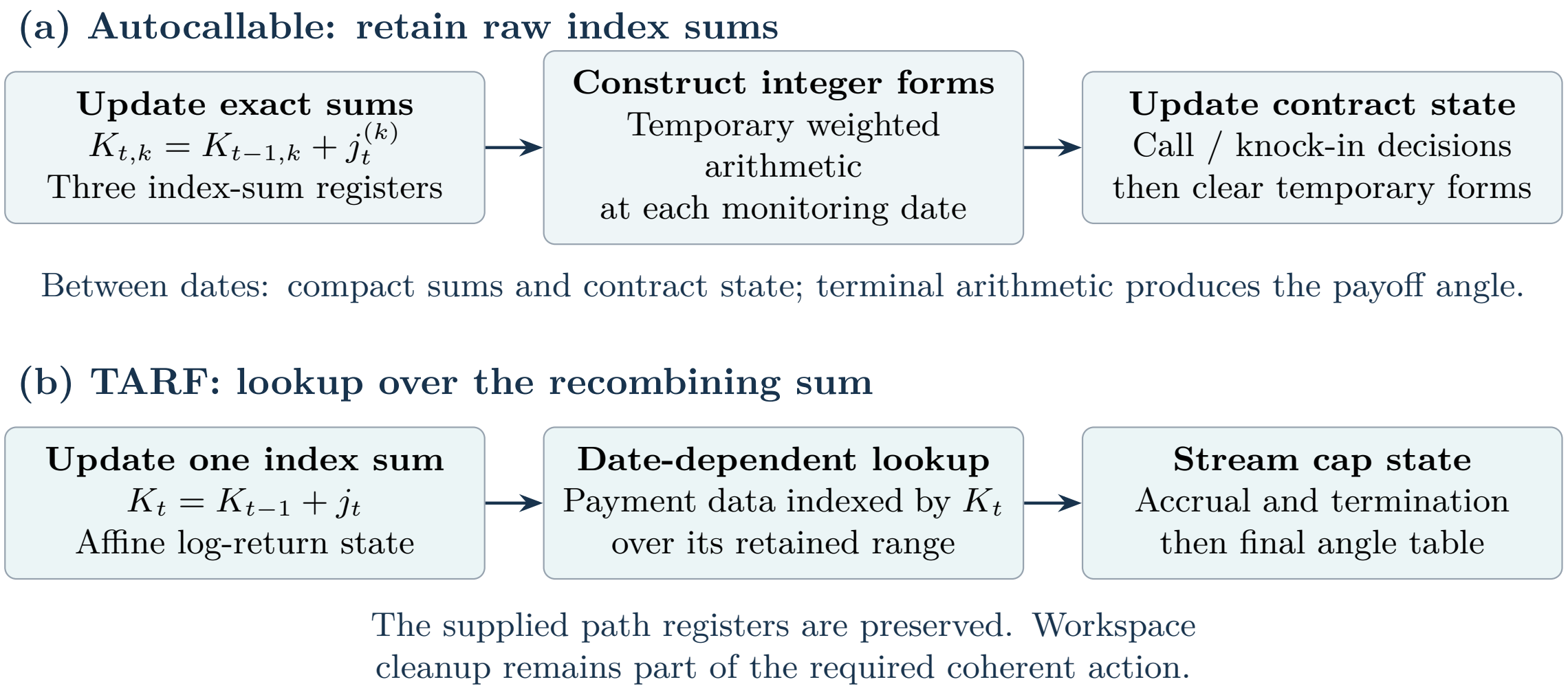
\begin{figure}[htbp]
\centering
\resizebox{\linewidth}{!}{%
\begin{tikzpicture}
\node[note,anchor=west,font=\small\bfseries] at (-2.25,1.1) {(a) Autocallable: retain raw index sums};
\node[stage] (as) at (0,0) {\textbf{Update exact sums}\\$K_{t,k}=K_{t-1,k}+j_t^{(k)}$\\Three index-sum registers};
\node[stage] (ar) at (5,0) {\textbf{Construct integer forms}\\Temporary weighted arithmetic\\at each monitoring date};
\node[stage] (ad) at (10,0) {\textbf{Update contract state}\\Call / knock-in decisions\\then clear temporary forms};
\draw[flow] (as)--(ar);\draw[flow] (ar)--(ad);
\node[note,text width=13.8cm] at (5,-1.3) {Between dates: compact sums and contract state; terminal arithmetic produces the payoff angle.};
\node[note,anchor=west,font=\small\bfseries] at (-2.25,-2.2) {(b) TARF: lookup over the recombining sum};
\node[stage,fill=Sea!8] (ts) at (0,-3.3) {\textbf{Update one index sum}\\$K_t=K_{t-1}+j_t$\\Affine log-return state};
\node[stage,fill=Sea!8] (tl) at (5,-3.3) {\textbf{Date-dependent lookup}\\Payment data indexed by $K_t$\\over its retained range};
\node[stage,fill=Sea!8] (tc) at (10,-3.3) {\textbf{Stream cap state}\\Accrual and termination\\then final angle table};
\draw[flow] (ts)--(tl);\draw[flow] (tl)--(tc);
\node[note,text width=13.8cm] at (5,-4.65) {The supplied path registers are preserved. Workspace cleanup remains part of the required coherent action.};
\end{tikzpicture}}
\caption{Retained state and temporary arithmetic in the two finance constructions. Arrows show the datewise data flow; contract history persists alongside the index sums. Boxes denote arithmetic stages rather than primitive gates.}
\label{fig:representations}
\end{figure}

Each construction chooses what to retain between dates and what to reconstruct temporarily. For each contract we first describe two reference circuits built from published approaches and then the compact circuits. Figure~\ref{fig:representations} summarizes the retained state of the main constructions.

\subsection{Autocallable}
The autocallable needs three asset states, a knock-in record and call records. The reference circuits store asset states as prices or as fixed-point log-returns. The compact circuits store exact integer index sums or barrier-centred integer forms, which are narrower and recombine across paths.

\subsubsection{Reference circuits}
The \emph{price-space reconstruction} is our reconstruction of Algorithm~5.1 of \citet{chakrabarti} under the common digital-angle interface: whereas Algorithm~5.1 assumes cumulative returns are already available from path loading and uses direct rotations for call payoffs, our reconstruction starts from factor-index inputs, charges the price conversion to the measured payoff-arithmetic component, and writes a 22-bit angle word. At every date each cumulative log-return is mapped to a price with a piecewise-cubic exponential~\citep{hrs}, the worst-of is taken on prices, and the put angle is evaluated from the worst price. To keep the comparison conservative, it shares the exact path lookups, measurement-uncomputed adders, streaming schedule, breach counter and fused angle table of our log-return reference, and it keeps the last price set live instead of recomputing it.

The \emph{log-return reference} applies log-domain comparisons~\citep{wang}. It retains one fixed-point log-return per asset, with 26 fractional bits and an offset that places the barrier at zero, a five-bit breach counter and call records. Five-bit table lookups supply correlated increments, sign tests update the breach counter, and a piecewise-linear table evaluates the terminal put angle. This avoids every exponential; the price-space reconstruction has a 523 times larger $\TQ$.

\subsubsection{Factor-sum circuit}
The factor-sum circuit replaces rounded log-return registers by exact index sums. With $\gamma=10/32$ and $z_j=z_0+\gamma j$, Eq.~\eqref{eq:autopath} becomes
\begin{equation}
\begin{gathered}
K_{t,k}=\sum_{s=1}^{t}j_s^{(k)},\qquad
C_i(t)=t d_i+\sum_{k=0}^{2}b_{ik}K_{t,k},\\
d_i=(r-\sigma_i^2/2)\Delta t+z_0\sum_{k=0}^{2}L_{ik},\qquad
b_{ik}=\gamma L_{ik},
\end{gathered}
\label{eq:index-affine}
\end{equation}
with per-date constants $d_i$ and $z_0=-5+\gamma/2$. Since $0\le K_{t,k}\le620$, three ten-bit registers hold the market state without wrap. A knock-in counter and two-bit call counters retain the history.

For the original parameters, $b_{11}/b_{10}=\sqrt3$ and $b_{22}/b_{20}=\sqrt{8/3}$. The circuit therefore compares integer forms
\begin{equation}
F_0=K_{t,0},\qquad F_1=qK_{t,0}+pK_{t,1},\qquad F_2=3qK_{t,0}+pK_{t,1}+r_fK_{t,2},
\label{eq:integer-forms}
\end{equation}
with $q=4128$, $p=7150$ and $r_f=20223$, against date-dependent integer thresholds for the barrier and call levels. Signed digits express each coefficient as shifted copies of the sum registers, and column reduction~\citep{wallace} keeps sum bits as parities of existing qubits. Only nonlinear carries need temporary qubits, so each form is a \emph{virtual register} whose top carry is the threshold decision; measurement clears the carries without further Toffoli gates (Appendix~\ref{app:virtual}). Asset 0 is tested for knock-in only from date 12, a deliberate approximation. At maturity a signed worst-of selection feeds a prefix-tree put table with Booth product rows, and a reverse sweep restores the sums and history.

Relative to the log-return reference, the circuit uses 55\% fewer Toffoli gates and 7\% fewer peak qubits. The saving accompanies approximation: the coefficient ratios $p/q$ and $r_f/q$ differ from $\sqrt3$ and $\sqrt{24}$ by $2.3\times10^{-5}$ and $3.1\times10^{-6}$, and terminal truncation, finite range and product rounding add further error. Section~\ref{sec:verification} separates the exact integer action from this financial error.

\subsubsection{Barrier-centred circuits}
A second construction retains, for each asset, the signed difference $D_i(t)$ between its integer form and the knock-in threshold, updated by date-dependent constants. The sign bit of each difference is the barrier test, so all three assets are tested at every date without separate comparators, and call tests become offset comparisons. The \emph{hardened circuit} uses finer forms, with $(q,p)=(1922,3329)$ for asset 1 and coefficients $(100336,57929,163848)$ for asset 2, and register widths that cannot wrap for any of the $32^{60}$ inputs. Its terminal stage converts the differences to a common unit with exact or five-digit constants and evaluates a minimax-fitted put table with Booth rows. The \emph{lowest-product circuit} keeps a closely related state but narrows widths where overflow is rare under the loading law, defers register widening, merges the terminal payoff into one table stage and tunes decision constants on sampled paths. Relative to factor-sum, the hardened circuit uses 9\% fewer Toffoli gates and 9\% fewer qubits while reducing selected threshold violations from 241 to 6, whereas the lowest-product circuit uses 33\% fewer Toffoli gates and 15\% fewer qubits at the cost of higher threshold sensitivity (Section~\ref{sec:robustness}). Appendix~\ref{app:other-circuits} gives further details, and Appendix~\ref{app:fixed-control} describes a fixed-point control.

\subsection{TARF}
TARF has one factor but a path-dependent accrual and cap. Again two reference circuits come from published approaches, and the compact circuits narrow the retained index and accrual words.

\subsubsection{Reference circuits}
The \emph{price-space reconstruction} follows Algorithm~5.2 of \citet{chakrabarti}: each date computes $S_t$ with a piecewise-cubic exponential, compares it with the knock-out and strike levels, forms the payment by arithmetic and discounts it by a constant multiplication. The \emph{lookup reference} instead retains the index sum $K_t=\sum_{s\le t}j_s$ and looks up each date's payment and discounted value, as published payoff operators do with exponential tables~\citep{qsharp}. Tabulation removes the coherent exponential and lowers $\TQ$ 199.5-fold.

\subsubsection{Windowed circuits}
With $P_{\mathrm{full}}=\sum_t h_t$ the sum of full uncapped payments up to termination and $Y=\sum_t(d_t-1)h_t$, a cap reached at date $\tau$ with excess $E=P_{\mathrm{full}}-5$ gives
\begin{equation}
f_{\mathrm{disc}}=5+Y+(1-d_\tau)E,
\label{eq:tarf-cap-correction}
\end{equation}
and $f_{\mathrm{disc}}=P_{\mathrm{full}}+Y$ without a cap. Since $|Y|<11$, a circuit can retain the shifted accrual $T_p=P_{\mathrm{full}}-5$ and the narrow correction $Y$ instead of a wide discounted total, and apply one cap-date product. The \emph{windowed circuit} keeps seven cyclic index bits over a date-dependent window, $T_p$ and $Y$ with 15 fractional bits, and alive and stop-date records. A prefix-tree lookup supplies the payment, the discount correction and a knock-out bit, and the sign of $T_p$ tests the cap. After the loop, a Booth-recoded cap product and an integer square root feed a piecewise-linear angle table.

The \emph{lowest-product circuit} narrows the date words further, with date-dependent widths and a lower window chosen by an absorbing-chain calculation, and simplifies the cap product, square root and lookup cleanup. The \emph{cap-hardened circuit} accumulates $T_p$ with 22 fractional bits, which removes selected cap-tie failures at a small cost. In all windowed circuits, a path below the retained window can alias a different payment, and finite-precision accrual can move the cap date (Appendix~\ref{app:tarf-exceptions}).

\subsection{Second parameter setting}
\label{sec:latticeforms}
When volatilities and correlation change, the compact coefficient ratios are lost; a generic factor-sum transfer has 5.58 times the $\TQ$ of the transferred log-return reference. Setting-specific lattice-form circuits instead accumulate fitted integer forms $Y_i(t)=\sum_kW_{ik}K_{t,k}$ directly: a shift-add circuit, a lookup circuit with two 32-entry tables, and a barrier-centred circuit with virtual-register forms and constants fitted to this setting (Appendix~\ref{app:second-parameters}).

\section{Design process and resource results}
\label{sec:search-results}
This section describes how the circuits were developed and compares their resources with the references of Section~\ref{sec:constructions}.

\subsection{Evaluator-guided design}
\label{sec:process}
Following \citet{gu}, the design process separates architecture proposals, executable design and evaluation. The authors fixed each contract, the interface, the accuracy and resource rules, the permitted access and the selection rule. In budgeted search sessions, an AI agent proposed architecture families from the task specification and builder interface, then implemented a family as a Python builder program with fitting or emulator helpers and revised it from development counts, errors and counterexamples. A later optimization stage used coordinated agents with access to all project artifacts, including the retained circuits, proofs and selected stress inputs, and fitted constants on sampled loading-law paths.

Executing a builder program emits a fixed gate tape. A separate-process scorer validates the tape, recounts resources and simulates basis values, phases and measurement corrections; all reported counts come from it. A candidate \emph{passes empirical screening} if, in $n=10^5$ sampled paths, at most $\lfloor n\varepsilon_{\mathrm{fail}}\rfloor$ angle-tolerance violations and no input, workspace or phase failures occur. Frozen circuits then received separate holdouts, independent-reference studies and, where stated, exact proofs (Section~\ref{sec:verification}). References and controls were prepared separately. Appendix~\ref{app:search-protocols} and Figure~\ref{fig:design-workflow} record the roles, budgets, access and deviations; the AI systems are listed at the end of the paper.

\subsection{Resource results}
\label{sec:resources}
\begin{figure}[p]\centering
\ref{finlegend}\par\medskip
\begin{tikzpicture}
\begin{semilogxaxis}[paperaxis,axis on top=false,width=0.94\linewidth,height=5.8cm,log ticks with fixed point,xticklabel style={/pgf/number format/1000 sep={\,}},title style={font=\small\bfseries},xlabel={Toffoli gates $N_{\mathrm{Tof}}$},ylabel={Peak live qubits $Q_{\max}$},grid=major,major grid style={Ink!8},clip mode=individual,title={(a) Basket autocallable},
xmin=4000,xmax=26000,ymin=425,ymax=640,xtick={5000,6000,8000,10000,15000,20000,25000},
legend to name=finlegend,legend columns=4,legend style={font=\footnotesize,draw=Ink!25,inner sep=3pt,/tikz/every even column/.append style={column sep=10pt}}]
\addplot[Ink!30,dashed,thin,smooth,forget plot] coordinates {(14603.7,640.00) (15037.0,621.56) (15483.2,603.64) (15942.7,586.25) (16415.7,569.35) (16902.8,552.95) (17404.4,537.01) (17920.8,521.54) (18452.6,506.51) (19000.1,491.91) (19563.9,477.73) (20144.5,463.97) (20742.2,450.60) (21357.7,437.61) (21991.4,425.00)};
\addplot[Ink!30,dashed,thin,smooth,forget plot] coordinates {(6111.6,640.00) (6292.9,621.56) (6479.7,603.64) (6671.9,586.25) (6869.9,569.35) (7073.8,552.95) (7283.7,537.01) (7499.8,521.54) (7722.3,506.51) (7951.5,491.91) (8187.4,477.73) (8430.4,463.97) (8680.5,450.60) (8938.1,437.61) (9203.3,425.00)};
\addplot[Ink!30,dashed,thin,smooth,forget plot] coordinates {(4000.0,560.79) (4080.0,549.79) (4161.6,539.01) (4244.8,528.44) (4329.7,518.08) (4416.3,507.92) (4504.7,497.96) (4594.8,488.20) (4686.7,478.62) (4780.4,469.24) (4876.0,460.04) (4973.6,451.01) (5073.0,442.17) (5174.5,433.50) (5278.0,425.00)};
\addplot[only marks,mark=square*,mark size=2.6pt,Ink] coordinates {(16810,556)};
\addlegendentry{Reference}
\addplot[only marks,mark=triangle*,mark size=3.2pt,Ink!60] coordinates {(19074,610)};
\addlegendentry{Control}
\addplot[only marks,mark=*,mark size=3.0pt,Sea] coordinates {(7551,518) (6888,472) (5075,442)};
\addlegendentry{Proposed circuit}
\addplot[only marks,mark=*,mark size=2.1pt,Sea,fill=white,thick] coordinates {(8521,482) (9487,477) (8641,474)};
\addlegendentry{Screened design candidate}
\node[font=\footnotesize,fill=white,fill opacity=0.9,text opacity=1,inner sep=1.5pt,anchor=north east,text=Ink,xshift=-2pt,yshift=-2pt] at (axis cs:16810,556) {Log-return reference};
\node[font=\footnotesize,fill=white,fill opacity=0.9,text opacity=1,inner sep=1.5pt,anchor=south,text=Ink,yshift=4pt] at (axis cs:19074,610) {Fixed-point control};
\node[font=\footnotesize,fill=white,fill opacity=0.9,text opacity=1,inner sep=1.5pt,anchor=north east,text=Sea,xshift=-3pt,yshift=-2pt] at (axis cs:7551,518) {Factor-sum, $R=2.39$};
\node[font=\footnotesize,fill=white,fill opacity=0.9,text opacity=1,inner sep=1.5pt,anchor=east,text=Sea,xshift=-6pt] at (axis cs:6888,472) {Hardened, $R=2.87$};
\node[font=\footnotesize,fill=white,fill opacity=0.9,text opacity=1,inner sep=1.5pt,anchor=west,text=Sea,xshift=6pt] at (axis cs:5075,442) {Lowest product, $R=4.17$};
\end{semilogxaxis}
\end{tikzpicture}
\par\medskip
\begin{tikzpicture}
\begin{semilogxaxis}[paperaxis,axis on top=false,width=0.94\linewidth,height=5.8cm,log ticks with fixed point,xticklabel style={/pgf/number format/1000 sep={\,}},title style={font=\small\bfseries},xlabel={Toffoli gates $N_{\mathrm{Tof}}$},ylabel={Peak live qubits $Q_{\max}$},grid=major,major grid style={Ink!8},clip mode=individual,title={(b) TARF},
xmin=3000,xmax=40000,ymin=235,ymax=400,xtick={3000,4000,5000,6000,8000,10000,15000,20000,30000,40000}]
\addplot[Ink!30,dashed,thin,smooth,forget plot] coordinates {(27290.9,400.00) (28046.4,389.22) (28822.9,378.74) (29620.9,368.54) (30441.0,358.61) (31283.7,348.95) (32149.8,339.55) (33039.9,330.40) (33954.6,321.50) (34894.7,312.84) (35860.7,304.41) (36853.6,296.21) (37873.9,288.23) (38922.4,280.46) (40000.0,272.91)};
\addplot[only marks,mark=square*,mark size=2.6pt,Ink] coordinates {(28428,384)};
\addplot[only marks,mark=*,mark size=2.1pt,Sea,fill=white,thick] coordinates {(10850,297) (11805,301) (10040,286) (9874,280) (8894,270)};
\addplot[only marks,mark=*,mark size=3.0pt,Sea] coordinates {(8381,269) (7553,262) (7921,277)};
\node[font=\footnotesize,fill=white,fill opacity=0.9,text opacity=1,inner sep=1.5pt,anchor=north east,text=Ink,xshift=-2pt,yshift=-2pt] at (axis cs:28428,384) {Lookup reference};
\node[font=\footnotesize,fill=white,fill opacity=0.9,text opacity=1,inner sep=1.5pt,anchor=north west,text=Sea,xshift=2pt,yshift=-5pt] at (axis cs:8381,269) {Windowed, $R=4.84$};
\node[font=\footnotesize,fill=white,fill opacity=0.9,text opacity=1,inner sep=1.5pt,anchor=east,text=Sea,xshift=-6pt] at (axis cs:7553,262) {Lowest product, $R=5.52$};
\node[font=\footnotesize,fill=white,fill opacity=0.9,text opacity=1,inner sep=1.5pt,anchor=east,text=Sea,xshift=-6pt] at (axis cs:7921,277) {Cap-hardened, $R=4.98$};
\node[font=\scriptsize,text=Ink!65,anchor=north west,inner sep=2pt] at (axis cs:3100,397) {Price-space reconstruction~\citep{chakrabarti} off scale ($3.33\times10^{6}$ Toffolis, 653 qubits)};
\end{semilogxaxis}
\end{tikzpicture}
\par\medskip
\begin{tikzpicture}
\begin{semilogxaxis}[paperaxis,axis on top=false,width=0.94\linewidth,height=5.8cm,log ticks with fixed point,xticklabel style={/pgf/number format/1000 sep={\,}},title style={font=\small\bfseries},xlabel={Toffoli gates $N_{\mathrm{Tof}}$},ylabel={Peak live qubits $Q_{\max}$},grid=major,major grid style={Ink!8},clip mode=individual,title={(c) Autocallable, second parameter setting},
xmin=2500,xmax=150000,ymin=420,ymax=680,xtick={3000,5000,10000,20000,50000,100000}]
\addplot[Ink!30,dashed,thin,smooth,forget plot] coordinates {(13744.6,680.00) (14225.9,656.99) (14724.1,634.77) (15239.6,613.29) (15773.3,592.54) (16325.6,572.50) (16897.3,553.13) (17488.9,534.42) (18101.3,516.34) (18735.2,498.87) (19391.2,481.99) (20070.2,465.68) (20773.0,449.93) (21500.4,434.71) (22253.2,420.00)};
\addplot[Ink!30,dashed,thin,smooth,forget plot] coordinates {(76754.8,680.00) (79442.5,656.99) (82224.2,634.77) (85103.4,613.29) (88083.4,592.54) (91167.8,572.50) (94360.1,553.13) (97664.2,534.42) (101084.0,516.34) (104623.6,498.87) (108287.1,481.99) (112078.9,465.68) (116003.5,449.93) (120065.5,434.71) (124269.7,420.00)};
\addplot[Ink!30,dashed,thin,smooth,forget plot] coordinates {(3206.3,680.00) (3318.6,656.99) (3434.8,634.77) (3555.1,613.29) (3679.6,592.54) (3808.4,572.50) (3941.8,553.13) (4079.8,534.42) (4222.6,516.34) (4370.5,498.87) (4523.5,481.99) (4681.9,465.68) (4845.9,449.93) (5015.6,434.71) (5191.2,420.00)};
\addplot[only marks,mark=square*,mark size=2.6pt,Ink] coordinates {(16810,556) (81552,640)};
\addplot[only marks,mark=triangle*,mark size=3.2pt,Ink!60] coordinates {(18822,610)};
\addplot[only marks,mark=*,mark size=3.0pt,Sea] coordinates {(12685,547) (13661,524) (4944,441)};
\node[font=\footnotesize,fill=white,fill opacity=0.9,text opacity=1,inner sep=1.5pt,anchor=south west,text=Ink,xshift=3pt,yshift=2pt] at (axis cs:16810,556) {Transferred reference};
\node[font=\footnotesize,fill=white,fill opacity=0.9,text opacity=1,inner sep=1.5pt,anchor=west,text=Ink,xshift=5pt] at (axis cs:18822,610) {Fixed-point control};
\node[font=\footnotesize,fill=white,fill opacity=0.9,text opacity=1,inner sep=1.5pt,anchor=south east,text=Ink,xshift=-8pt,yshift=2pt] at (axis cs:81552,640) {Generic factor-sum transfer, $R=0.18$};
\node[font=\footnotesize,fill=white,fill opacity=0.9,text opacity=1,inner sep=1.5pt,anchor=east,text=Sea,xshift=-5pt] at (axis cs:12685,547) {Shift-add, $R=1.35$};
\node[font=\footnotesize,fill=white,fill opacity=0.9,text opacity=1,inner sep=1.5pt,anchor=east,text=Sea,xshift=-5pt] at (axis cs:13661,524) {Lookup lattice form, $R=1.31$};
\node[font=\footnotesize,fill=white,fill opacity=0.9,text opacity=1,inner sep=1.5pt,anchor=west,text=Sea,xshift=6pt] at (axis cs:4944,441) {Barrier-centred lattice form, $R=4.29$};
\end{semilogxaxis}
\end{tikzpicture}
\caption{Finance resource tradeoffs. Labels give $R_{\mathrm{L}}$ of Eq.~\eqref{eq:resource-ratio} against the log-return or lookup reference in each panel (Table~\ref{tab:resources}); $R<1$ means a higher product. Filled circles are the proposed circuits and hollow circles screened design candidates. Dashed curves have equal $\TQ$; position ranks resources, not accuracy. The price-space reconstructions of \citet{chakrabarti} lie off scale, with $\TQ$ of $4.89\times10^{9}$ (autocallable) and $2.18\times10^{9}$ (TARF). Variants with nearly equal counts are listed in Table~\ref{tab:variants}.}
\label{fig:resources}\end{figure}
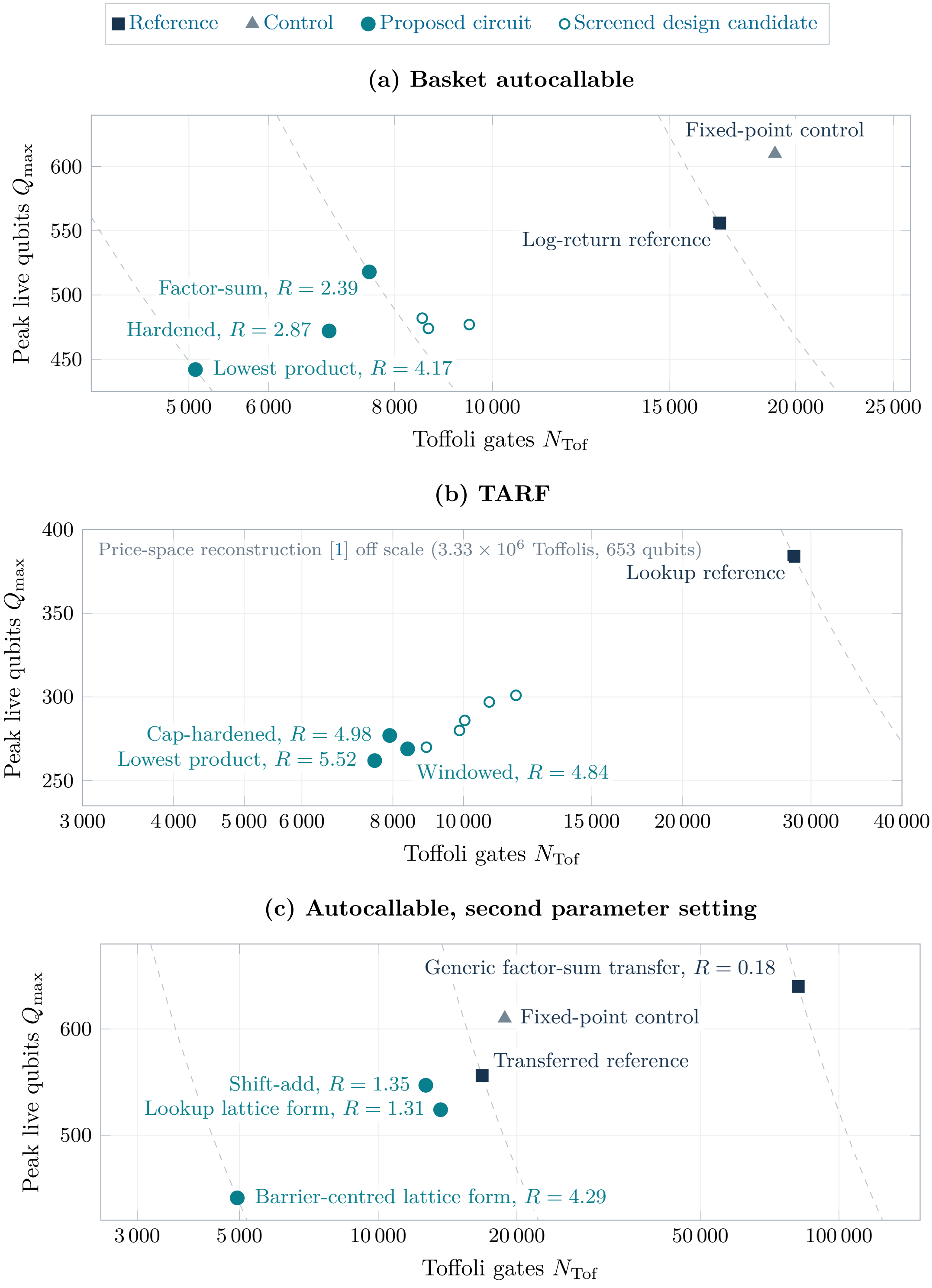

For a reference circuit and a candidate circuit, define
\begin{equation}
R=\frac{N_{\mathrm{Tof}}^{\mathrm{reference}}Q_{\max}^{\mathrm{reference}}}{N_{\mathrm{Tof}}^{\mathrm{candidate}}Q_{\max}^{\mathrm{candidate}}}.
\label{eq:resource-ratio}
\end{equation}
Table~\ref{tab:resources} gives each circuit's counts and its ratios $R_{\mathrm{P}}$ against the price-space reconstruction and $R_{\mathrm{L}}$ against the log-domain reference of the same contract. Figure~\ref{fig:resources} shows the resource planes.

\begin{table}[htbp]\centering\small
\caption{Secure payoff-arithmetic counts. $R_{\mathrm{P}}$: ratio of Eq.~\eqref{eq:resource-ratio} against our price-space reconstructions of \citet{chakrabarti} under the common digital-angle interface; $R_{\mathrm{L}}$: against the log-return (autocallable) or lookup (TARF) reference. The second setting has no price-space reconstruction. Ratios compare the payoff-arithmetic component only.}\label{tab:resources}
\setlength{\tabcolsep}{6pt}\renewcommand{\arraystretch}{1.08}
\begin{tabularx}{\linewidth}{@{}>{\raggedright\arraybackslash}Xrrrr@{}}\toprule
Circuit & $N_{\mathrm{Tof}}$ & $Q_{\max}$ & $R_{\mathrm{P}}$ & $R_{\mathrm{L}}$\\\midrule
\multicolumn{5}{@{}l}{\emph{Autocallable, original setting}}\\\addlinespace[2pt]
Price-space reconstruction~\citep{chakrabarti} & 5,433,065 & 900 & 1 & 0.0019\\
Log-return reference & 16,810 & 556 & 523 & 1\\
Fixed-point control & 19,074 & 610 & 420 & 0.80\\
Factor-sum circuit & 7,551 & 518 & 1,250 & 2.39\\
Hardened circuit & 6,888 & 472 & 1,504 & 2.87\\
Lowest-product circuit & 5,075 & 442 & 2,180 & 4.17\\
\addlinespace[4pt]
\multicolumn{5}{@{}l}{\emph{TARF}}\\\addlinespace[2pt]
Price-space reconstruction~\citep{chakrabarti} & 3,334,652 & 653 & 1 & 0.0050\\
Lookup reference & 28,428 & 384 & 199.5 & 1\\
Windowed circuit & 8,381 & 269 & 966 & 4.84\\
Cap-hardened circuit & 7,921 & 277 & 992 & 4.98\\
Lowest-product circuit & 7,553 & 262 & 1,100 & 5.52\\
\addlinespace[4pt]
\multicolumn{5}{@{}l}{\emph{Autocallable, second setting}}\\\addlinespace[2pt]
Transferred log-return reference & 16,810 & 556 & -- & 1\\
Generic factor-sum transfer & 81,552 & 640 & -- & 0.18\\
Shift-add lattice-form circuit & 12,685 & 547 & -- & 1.35\\
Lookup lattice-form circuit & 13,661 & 524 & -- & 1.31\\
Barrier-centred lattice-form circuit & 4,944 & 441 & -- & 4.29\\\bottomrule
\end{tabularx}
\end{table}

Relative to the log-return and lookup references, which already avoid coherent exponentials, the ratios $R_{\mathrm{L}}$ isolate the gain from compact representations: 2.39 to 4.17 for the autocallable and 4.84 to 5.52 for TARF. Against our conservative price-space reconstructions of \citet{chakrabarti} under the common digital-angle interface (which start from factor-index inputs, charge price conversion to the measured component and write a 22-bit angle word, whereas Algorithm~5.1 assumes cumulative returns from path loading and uses direct rotations for call payoffs), the ratios $R_{\mathrm{P}}$ are 1,250 to 2,180 and 966 to 1,100; most of that larger gap comes from removing coherent exponentials. The fixed-point control, which uses conventional rounding and full-range widths, has a larger product than the log-return reference, so precision tuning alone does not reproduce the saving. The lowest-product autocallable is more sensitive to threshold inputs, whereas targeted hardening improves selected boundary cases while preserving resource savings (Section~\ref{sec:robustness}); Figure~\ref{fig:resources} ranks resources, not accuracy. Figure~\ref{fig:profile} in Appendix~\ref{app:stage-accounting} locates most autocallable savings in the date sweeps.

\subsection{Parameter transfer and scope}
\label{sec:parameter-transfer}
At the second setting, the fixed factor-sum transfer loses its advantage ($R_{\mathrm{L}}=0.18$), while setting-specific lattice-form circuits recover 1.31- to 4.29-fold savings. The best of these is close to the lowest product at the original setting, but its constants are fitted to this setting, so each parameter set needs its own fitted circuit. One further setting does not establish general transfer or scaling with dates, assets or precision. Published estimates for complete pricing~\citep{qsharp,qsp,integration} count different operations and accuracy boundaries; in particular, quantum signal processing combines payoff evaluation with amplitude encoding, whereas our component writes a digital angle.

\FloatBarrier

\section{Verification and pricing error}
\label{sec:verification}
\label{sec:price-integration}
Resource efficiency does not establish correctness or pricing accuracy. Compact integer representations, finite precision and truncated lookup ranges can make the implemented angle differ from the contract, and empirical screening does not guarantee correct coherent behaviour on all inputs. We therefore separate the exact implementation of each integer specification from its numerical agreement with the financial payoff.

\subsection{Exact logical action and inverse}
\label{sec:structural}
\label{sec:outputproof}
We use the evaluator's semantics: $X$, CNOT, Toffoli, SWAP and Fredkin act as reversible Boolean gates; $Z$, CZ, CCZ and global signs supply phases, and only these diagonal gates may be conditioned on earlier measurement bits. Allocation supplies $|0\rangle$, freeing is permitted only at zero, and the sole measurement, \texttt{hmr}, measures an ancilla in the $X$ basis and resets it. For a frozen tape $\mathcal T$, let $\Delta_{\mathcal T}$ be its integer model (Appendix~\ref{app:integer-model} for the factor-sum tape, and Appendix~\ref{app:other-circuits} for the hardened tape) and $U_\Delta$ the clean-workspace modular addition of Eq.~\eqref{eq:oracle} with $\widehat A=\Delta_{\mathcal T}$.

\begin{theorem}[Coherent integer action]
\label{thm:factor-sum}
Let $\mathcal T$ be the frozen factor-sum tape (259,386 operations, $M=7,437$ measurements) or the frozen hardened tape (181,465 operations, $M=6,556$ measurements). For every $\mathbf z\in\{0,\ldots,31\}^{60}$ and every $\theta\in\mathbb Z/2^{22}\mathbb Z$, with workspace initialized to zero, the corrected implementation of $\mathcal T$ realizes
\begin{equation}
U_\Delta|\mathbf z\rangle|\theta\rangle|0\rangle
=|\mathbf z\rangle|\theta+\Delta_{\mathcal T}(\mathbf z)\bmod2^{22}\rangle|0\rangle.
\label{eq:coherent-factor-sum}
\end{equation}
Every input is preserved, every workspace qubit is restored to zero before release, and unconditioned and measurement-dependent phases cancel. For each measurement record $\mathbf m$, the branch operator is $K_{\mathbf m}=2^{-M/2}U_\Delta$, so the record is uniform and independent of the input, and the action holds coherently on superpositions.
\end{theorem}

The proofs cut each tape between dates, where only narrow integer state and contract history persist. Exhaustive bitsliced evaluation proves the finite-domain value claims, and canonical XOR--AND cancellation proves that other live slots are unchanged and internal frees are clean. Each update maps its declared input range into the next segment's domain, so the local claims compose from zero initialization. Phase obligations close by cancellation, exhaustive evaluation of their support or certified unsatisfiability. These are computational certificates, not proof-assistant formalizations; Appendix~\ref{app:proof-records} gives the lemma inventories and trusted components.

\begin{corollary}[Logical inverse]
\label{cor:inverse}
With $J_m|\theta\rangle=|2^m-1-\theta\rangle$ on an $m$-bit accumulator, the actions of Theorem~\ref{thm:factor-sum} ($m=22$) and of the proved HHL circuits ($m=33$, Appendix~\ref{app:hhl}) satisfy $J_mU_\Delta J_m=U_\Delta^\dagger$ on clean workspace.
\end{corollary}
The inverse therefore has the same Toffoli count and peak qubits, adding only $X$ gates (Appendix~\ref{app:inverse-check}). As a calibration, the HHL angle circuits of Appendix~\ref{app:hhl} have the same kind of all-input logical proof and an exhaustive numerical accuracy check.

\subsection{Numerical financial error}
\label{sec:exact}
For a proved integer map, the violation probability is
\begin{equation}
p_{\mathrm{fail}}=\Pr_{\mathbf z\sim p_{\mathrm{load}}}
\left[|2^{-20}\Delta(\mathbf z)-a(\mathbf z)|>\delta_{\mathrm{ang}}\right].
\label{eq:pfail}
\end{equation}
Both the integer models and the financial target make their datewise tests on the index-sum vector, so a forward recursion over the lattice of index sums, with knock-in flags for both sides and absorbing call states, propagates the exact path law. At maturity the remaining mass is classified by evaluating the full integer terminal arithmetic against the financial angle. The calculation gives $p_{\mathrm{fail}}\approx\ExactFull{}$ for the factor-sum circuit and $3.56\times10^{-8}$ for the hardened circuit, with omitted mass below \ExactLost{} and $4.7\times10^{-13}$. For the hardened circuit, the numerical recursion found no decision mismatches on the retained lattice (with omitted mass bounded by $4.7\times10^{-13}$), and all remaining violations lie in the far put tail. The recursion is exact in real arithmetic, but its evaluation is numerical, without a certified enclosure (Appendix~\ref{app:lattice-records}).

\subsection{Validation and robustness}
\label{sec:independent-validation}
\label{sec:robustness}
The validation population is the loading law with a uniform incoming accumulator and the evaluator's measurement outcomes. For one circuit fixed before $n$ independent draws, zero violations give the one-sided upper bound $1-\alpha^{1/n}$ at confidence $1-\alpha$; at 95\%, zero in $10^6$ gives $2.99573\times10^{-6}$. Table~\ref{tab:evidence} collects the evidence. The factor-sum autocallable, the windowed TARF circuit, and the second-setting shift-add and lookup lattice-form circuits (together with the log-return and transferred log-return references) have zero violations in a million-path task holdout and in a separate million-path study against an independent 60/90-digit reference. The barrier-centred, cap-hardened and lowest-product circuits have one million-path holdout each, at a recorded seed (Appendix~\ref{app:statistics}). Every sampled study had zero input, phase and workspace errors.

\begin{table}[htbp]\centering\small
\caption{Validation evidence. \emph{Holdout}: angle-tolerance violations in $10^6$ task-reference paths; \emph{Indep.}: violations in $10^6$ paths against an independent 60/90-digit reference. \emph{Selected}: violations on retained stress inputs (900 threshold paths, 400 TARF cap-tie paths, 66 extreme second-setting paths), which are diagnostics, not failure rates. \emph{Proof}: Theorem~\ref{thm:factor-sum}. \textsuperscript{a}Single recorded seed; inputs were visible during design (Appendix~\ref{app:search-protocols}). \textsuperscript{b}100,000-path empirical screening only; no $10^6$-path holdout.}\label{tab:evidence}
\setlength{\tabcolsep}{5pt}\renewcommand{\arraystretch}{1.05}
\begin{tabularx}{\linewidth}{@{}>{\raggedright\arraybackslash}Xrrrc@{}}\toprule
Circuit & Holdout & Indep. & Selected & Proof\\\midrule
\multicolumn{5}{@{}l}{\emph{Autocallable, original setting}}\\\addlinespace[2pt]
Price-space reconstruction & 0/100k\textsuperscript{b} & -- & 52 & --\\
Log-return reference & 0 & 0 & 52 & --\\
Factor-sum circuit & 0 & 0 & 241 & yes\\
Hardened circuit & 0 & -- & 6\textsuperscript{a} & yes\\
Lowest-product circuit & 4 & -- & 520\textsuperscript{a} & --\\
\addlinespace[3pt]
\multicolumn{5}{@{}l}{\emph{TARF}}\\\addlinespace[2pt]
Price-space reconstruction & 0/100k\textsuperscript{b} & -- & -- & --\\
Lookup reference & 0 & -- & 2 & --\\
Windowed circuit & 0 & 0 & 47 & --\\
Cap-hardened circuit & 4 & -- & 0\textsuperscript{a} & --\\
Lowest-product circuit & 2 & -- & 33\textsuperscript{a} & --\\
\addlinespace[3pt]
\multicolumn{5}{@{}l}{\emph{Autocallable, second setting}}\\\addlinespace[2pt]
Transferred log-return reference & 0 & 0 & 26 & --\\
Shift-add / lookup lattice forms & 0 / 0 & 0 / 0 & 49 / 46 & --\\
Barrier-centred lattice forms & 1 & -- & 55\textsuperscript{a} & --\\\bottomrule
\end{tabularx}
\end{table}

Selected stress inputs show a tradeoff between cost and robustness. The factor-sum circuit fails 241 of 900 threshold paths, through coefficient rounding, omitted early asset-0 tests and terminal wrap. The hardened circuit tests every asset at every date and cannot wrap; it fails 6, all near the barrier. The lowest-product circuit fails 520 and has four holdout violations with decision-sized errors. For TARF, the cap-hardened circuit removes all selected cap-tie failures, while every windowed circuit fails selected paths below its window (Appendix~\ref{app:tarf-exceptions}). At the second setting, the barrier-centred circuit fails more extreme inputs than the transferred reference. For each of the two circuits with four violations in $10^6$ paths (the lowest-product autocallable and cap-hardened TARF), the per-circuit one-sided 95\% upper rate is $9.15\times10^{-6}$, inside but close to the $10^{-5}$ budget.

\subsection{Pricing error and integration}
\label{sec:price}
\label{sec:beyond}
Equation~\eqref{eq:effective-value} gives $|\widehat V_{\mathrm{grid}}-V_{\mathrm{grid}}|=D\,|\mathbb E[\sin^2\widehat a-\sin^2a]|$. Since $|\mathrm d\sin^2a/\mathrm da|\le1$, paths within tolerance contribute payoff error at most $\delta_{\mathrm{ang}}$ and violations at most one, so
\begin{equation}
\left|\mathbb E_{p_{\mathrm{load}}}[\sin^2\widehat a-\sin^2a]\right|\le\delta_{\mathrm{ang}}+p_{\mathrm{fail}}.
\label{eq:bias}
\end{equation}
For the factor-sum circuit, the numerical lattice value of the signed bias is \ExactBias{} in normalized units, or \ExactBiasPrice{} contract units; for the hardened circuit it is $+3.67\times10^{-7}$, or $+1.01\times10^{-5}$ contract units (Appendix~\ref{app:bias-diagnostic}). A complete pricing construction must add loading, payoff rotations, reflections and any controlled use required by the amplitude-estimation variant, with discretization, model, synthesis and estimation errors budgeted separately. Corollary~\ref{cor:inverse} supplies the logical inverse; controlled use and compiled costs remain open.

\section{Discussion and conclusion}
\label{sec:limits}
Compact retained state lowers cost and helps verification, because it limits what each date must update, reconstruct and restore. Index sums and barrier-centred forms keep the autocallable state narrow and recombining, which makes the all-input proofs and the lattice calculation of financial error tractable. TARF also carries path-dependent accrual and discount words, so its full error probability does not follow from the recombining index alone.

Most of the saving against price-space arithmetic comes from avoiding coherent exponentials; the further 2.4- to 5.5-fold savings come from the compact representations. These gains have limits. Lower products can trade robustness near contract thresholds for cost, and fitted constants tie a circuit to one parameter set: at the second setting, a fixed transfer loses its advantage while a setting-specific design recovers it. The design record also limits attribution: the AI-assisted runs had access to task files, some were assisted, and the circuits share design lineages (Appendix~\ref{app:search-protocols}). Repeated comparisons under matched budgets would be needed to assess the reliability of the design process.

The demonstrated result is a payoff-arithmetic component with quantified approximation error, not a complete pricing advantage. A complete comparison must align the loading law, accuracy budget, interface and compiled gate, workspace and feed-forward accounting. Within that scope, representation is a useful design variable for quantum components whose resource savings and approximation behaviour must be assessed together.

\section*{Author contributions and AI use}
\label{sec:contributions-ai}
Monit Sharma and Hoong Chuin Lau designed the study, fixed the tasks, evaluation rules and evidence standards, directed the circuit design, and reviewed all results and claims. The following AI systems were used. Claude Opus~5.5 (Anthropic) and Gemini~4 Neon (Google) served as agents during the project: Claude Opus~5.5 served as the agent in the budgeted search sessions and in the independent review of the optimization-stage circuits, whereas Gemini~4 Neon served in the coordinated multi-agent workflow that optimized the barrier-centred autocallable circuits, the narrower TARF circuits, the barrier-centred second-setting circuits and the signed-Booth HHL circuits and adapted the corresponding proof scripts; both models were used to write builder, fitting, emulator, verification and numerical-analysis code, rerun the evidence and assist with editing the manuscript. The authors take responsibility for the content and its claims.

\section*{Code availability}
The circuit programs, task specifications, gate-level scorer, verification and analysis scripts, computational records and figure files that support this study are openly available on Zenodo at \doi{10.5281/zenodo.24556531}.

\bibliographystyle{unsrtnat}
\bibliography{reference}
\appendix
\section{Integer models and arithmetic details}
\label{app:arithmetic-details}
\subsection{Payoff normalization and conventions}
\label{app:normalization}
For the autocallable, $H_{\mathrm{auto}}=18e^{-r}$ and $D_{\mathrm{auto}}=H_{\mathrm{auto}}+\max_{a=1,\ldots,5}2ae^{-ra/5}=27.44556285258915$. For TARF, each date pays at least $-40$, and capped partial payment sums with decreasing discount factors bound the total above by $5d_1$; hence $H_{\mathrm{tarf}}=40\sum_{t=1}^{26}d_t$ and $D_{\mathrm{tarf}}=H_{\mathrm{tarf}}+5d_1$, our declared normalization, which is tighter than the printed source bound. In both cases $\widetilde f=(f_{\mathrm{disc}}+H)/D\in[0,1]$ and $V_{\mathrm{grid}}=D\,\mathbb E_{p_{\mathrm{load}}}[\widetilde f]-H$. TARF comparisons include equality at the knock-out level and the cap; the algorithm example of \citet{chakrabarti} uses fifty-two dates, and we use their twenty-six-date resource benchmark. Reported calculations use the frozen binary64 loading-probability vector. The per-path tolerance $\delta_{\mathrm{ang}}$ is numerically ten times smaller than the payoff-precision budget $\epsilon_f=10^{-4}$ of \citet{chakrabarti}, but the two are defined differently and cannot be interchanged.

\subsection{Virtual-register example}
\label{app:virtual}
Let $X=x_0+2x_1$, $Y=y_0+2y_1$ and $Z=z_0+2z_1$ be three existing two-bit registers and consider $V=X+2Y+2Z$; all six lowercase variables are bits on existing qubits, and shifted copies are references to those qubits. With $\operatorname{maj}(a,b,c)$ the majority bit, column reduction gives:
\begin{center}\small
\begin{tabular}{@{}clll@{}}\toprule
Column & Incoming bits & Sum bit (parity) & Carry to next column\\\midrule
0 & $x_0$ & $s_0=x_0$ & none\\
1 & $x_1,y_0,z_0$ & $s_1=x_1\oplus y_0\oplus z_0$ & $g_2=\operatorname{maj}(x_1,y_0,z_0)$\\
2 & $y_1,z_1,g_2$ & $s_2=y_1\oplus z_1\oplus g_2$ & $g_3=\operatorname{maj}(y_1,z_1,g_2)$\\
3 & $g_3$ & $s_3=g_3$ & none\\\bottomrule
\end{tabular}
\end{center}
Then $V=\sum_{j=0}^3 2^js_j$, and only $g_2$ and $g_3$ need new qubits. A required parity is exposed on an existing qubit with CNOT gates, used and restored. The identity $\operatorname{maj}(a,b,c)=c\oplus[(a\oplus c)\land(b\oplus c)]$ computes each generic carry with one Toffoli gate. After the decision is used, the carries are cleared in reverse order by measurement; for outcome $m$, the phase $(-1)^{m\operatorname{maj}(a,b,c)}$ is cancelled by a conditional $Z$ on $c$ and a conditional CZ on the exposed parities. Cleanup therefore preserves superpositions without further Toffoli gates. The same reduction handles the longer signed-digit forms of Eq.~\eqref{eq:integer-forms}; a constant makes the top carry the threshold decision.

\subsection{Factor-sum integer model}
\label{app:integer-model}
The rules below specify the arithmetic and table interpretation for $\Delta$ in Theorem~\ref{thm:factor-sum}; the frozen factor-sum coefficient dataset identified by the artifact supplies its 292 indexed quadruples. Inputs are $j_t^{(k)}\in\{0,\ldots,31\}$, $t=1,\ldots,20$, $k=0,1,2$. Operations are integer except for the documented generation of fixed data. The literal thresholds, constants and frozen coefficients are authoritative; the generation formulas do not replace them with recomputed values.

\paragraph{1. State, initialization and date updates.}
Initialize three ten-bit sums $K_{0,k}=0$, a five-bit breach counter $H_0=0$ (the proof's $K_C$), and five two-bit call counters $C_a=0$ (the proof's $CC_a$), $a=0,\ldots,4$, for dates $s_a=4(a+1)$. At each date first update
\begin{equation}
 K_{t,k}=K_{t-1,k}+j_t^{(k)},\qquad 0\le K_{t,k}\le31t\le620.
 \label{eq:delta-sums}
\end{equation}
These additions do not wrap. Define the temporary forms
\begin{equation}
 \begin{gathered}
 F_0(t)=K_{t,0},\qquad F_1(t)=qK_{t,0}+pK_{t,1},\\
 F_2(t)=3qK_{t,0}+pK_{t,1}+r_fK_{t,2},\qquad
 (q,p,r_f)=(4128,7150,20223).
 \end{gathered}
 \label{eq:delta-forms}
\end{equation}
With $(n_0,n_1,n_2)=(1,q+p,3q+p+r_f)$, $0\le F_i(t)\le31tn_i$. Forms are temporary arithmetic expressions, not additional state retained between dates. Every date is processed, even after a call flag is set; the final output gives priority to the earliest call.

\paragraph{2. Fixed thresholds and constants.}
The literal thresholds $T_i^B(t),T_i^C(t)$ in Table~\ref{tab:delta-thresholds} and constants below define the frozen rules:
\begin{align*}
 (c_0,c_1,c_2)&=(-12721351,-49993977,-96494227),\\
 B&=158030,\qquad C_\varnothing=975463,\\
 (C_0^{\rm call},\ldots,C_4^{\rm call})&=(1057100,1144652,1243260,1365747,1647099).
\end{align*}
Thresholds and $B$ were generated with ceiling; offsets and angle words use numerical nearest-integer rounding with ties to even. The zero-payoff angle word $C_\varnothing$ is not zero. For the fitted target below, write $\ell_B=\ln0.7$, $\gamma=10/32$ and $u=\gamma L_{10}/(15q)\approx1.410633161677738\times10^{-7}$, using the original step Cholesky factor of Eq.~\eqref{eq:autopath}. The dataset retains the binary64 constants; its documentation gives their generation derivations. No recomputation replaces these literal integers or the fixed common-unit multipliers in Eq.~\eqref{eq:delta-terminal}.

\begin{table}[htbp]\centering\small\setlength{\tabcolsep}{5pt}
\caption{Literal integer thresholds of the frozen factor-sum model. Every date uses the asset-1/2 barrier columns; asset 0 is tested only from date 12. Call columns are used only at the five call dates. The other call-threshold entries are unused and shown as dashes.}\label{tab:delta-thresholds}
\begin{tabular}{rrrrrrr}\toprule
$t$ & $T_0^B(t)$ & $T_1^B(t)$ & $T_2^B(t)$ & $T_0^C(t)$ & $T_1^C(t)$ & $T_2^C(t)$\\\midrule
1 & -35 & 6510 & 302833 & --- & --- & ---\\
2 & -20 & 181585 & 921725 & --- & --- & ---\\
3 & -4 & 356660 & 1540617 & --- & --- & ---\\
4 & 11 & 531735 & 2159509 & 76 & 745344 & 2560025\\
5 & 26 & 706810 & 2778401 & --- & --- & ---\\
6 & 42 & 881885 & 3397293 & --- & --- & ---\\
7 & 57 & 1056959 & 4016184 & --- & --- & ---\\
8 & 73 & 1232034 & 4635076 & 137 & 1445643 & 5035592\\
9 & 88 & 1407109 & 5253968 & --- & --- & ---\\
10 & 103 & 1582184 & 5872860 & --- & --- & ---\\
11 & 119 & 1757259 & 6491752 & --- & --- & ---\\
12 & 134 & 1932334 & 7110644 & 199 & 2145942 & 7511160\\
13 & 150 & 2107408 & 7729536 & --- & --- & ---\\
14 & 165 & 2282483 & 8348428 & --- & --- & ---\\
15 & 180 & 2457558 & 8967319 & --- & --- & ---\\
16 & 196 & 2632633 & 9586211 & 260 & 2846242 & 9986728\\
17 & 211 & 2807708 & 10205103 & --- & --- & ---\\
18 & 227 & 2982783 & 10823995 & --- & --- & ---\\
19 & 242 & 3157857 & 11442887 & --- & --- & ---\\
20 & 257 & 3332932 & 12061779 & 322 & 3546541 & 12462295\\
\bottomrule\end{tabular}
\end{table}

\paragraph{3. Monitoring and call history.}
After updating the sums, set
\begin{align}
 e_{12}(t)&=\mathbf1\{F_1(t)<T_1^B(t)\ \text{or}\ F_2(t)<T_2^B(t)\},\nonumber\\
 e_0(t)&=\mathbf1\{t\ge12\}\mathbf1\{F_0(t)<T_0^B(t)\},\nonumber\\
 H_t&=H_{t-1}+e_{12}(t)+e_0(t),\qquad
 0\le H_t\le t+\max(0,t-11)\le29.
 \label{eq:delta-history}
\end{align}
Asset-1/2 breaches count once jointly per date; asset 0 counts separately only at dates 12--20. Its omitted tests at dates 1--11 are a deliberate approximation to the contract, not a claim that early breaches are impossible. At date $t=s_a$, update the initially zero counter to
\[
 C_a=\sum_{i=0}^{2}\mathbf1\{F_i(s_a)\ge T_i^C(s_a)\}\in\{0,1,2,3\};
 \qquad A_a=\mathbf1\{C_a=3\}.
\]
Other call counters remain unchanged. Breaches are strict, calls include equality, and the bounded history counters do not wrap. At maturity the knock-in flag is $h=\mathbf1\{H_{20}>0\}$ and the five-bit call pattern is $A=\sum_{a=0}^{4}2^aA_a$; these flags are derived from the retained counters.

\paragraph{4. Terminal words and signed wrap.}
Using the final sums and forms, compute
\begin{align}
 Z_0&=12qK_{20,0}+c_0,\qquad Z_1=15F_1(20)+c_1,\qquad Z_2=8F_2(20)+c_2,\nonumber\\
 v_i&=\left\lfloor\frac{Z_i\bmod2^{25}}{16}\right\rfloor\in\{0,\ldots,2^{21}-1\},\qquad
 Y_i=\begin{cases}v_i,&v_i<2^{20},\\v_i-2^{21},&v_i\ge2^{20},\end{cases}\nonumber\\
 y&=\min(Y_0,Y_1,Y_2)\in[-2^{20},2^{20}-1].
 \label{eq:delta-terminal}
\end{align}
Equivalently, $Y_i=((\lfloor Z_i/16\rfloor+2^{20})\bmod2^{21})-2^{20}$. Dropping four low bits means floor division, including for negative values; it is not rounding toward zero or nearest rounding. The words are interpreted as signed two's-complement integers before taking the minimum. Large terminal values wrap, rather than saturating or being clipped. A wrapped word can become the false worst asset, as in the retained counterexamples.

\paragraph{5. Put address, table data and fallback.}
The put branch is eligible only if $A=0$, $h=1$ and $y<B$; the last comparison is strict. Its unsigned address is $v=y+2^{20}\in\{0,\ldots,2^{21}-1\}$. The fixed table has 292 disjoint prefix cells, each stored as a quadruple $(\lambda,p,c,s)$ with depth $\lambda\in\{9,10\}$, prefix $0\le p<2^\lambda$, intercept $c$ and nonnegative slope word $s$. It selects $(c(v),s(v))=(c,s)$ when
\begin{equation}
 p2^{21-\lambda}\le v<(p+1)2^{21-\lambda}.
 \label{eq:delta-prefix}
\end{equation}
The cells here cover $139264\le v<1208320$. Addresses outside the covered cells return $C_\varnothing$ even when the put branch is eligible: there is no extrapolation or clipping to an endpoint. The stored intercepts lie in $[254579,971696]$, slopes in $[2408,13036]$. The circuit packs the low 21 bits as $c\mathbin{\oplus}C_\varnothing$ and the upper 14 bits as $s$, to XOR the put correction into the initially loaded no-call angle word. The mathematical definition below uses the resulting intercept $c$ directly.

\paragraph{6. Table target and numerical fit.}
Let $\delta_v=v\bmod2^{12}$, not the offset from the selected cell's left endpoint. The fitted target was
\[
 \Phi(v)=2^{20}\arcsin\sqrt{P_{\rm put}\exp[\ell_B+u(16(v-2^{20})+7.5)]},
 \qquad P_{\rm put}=18e^{-r}/D_f,
\]
where $D_f=D_{\mathrm{auto}}$ is the normalization denominator defined in Section~\ref{sec:task}. The $7.5$ is the midpoint of the sixteen values discarded from a terminal word. This target explains the coefficients; it is not substituted for their integer evaluation.

The fit evaluates every integer address in a cell using the implemented rounded row sum $R_s(v)$ of Eq.~\eqref{eq:delta-rows}, rather than an ideal real product. Its numerical acceptance criterion is not a certified contract-error bound. The frozen array determines the implemented function independently of refitting; the artifact documentation retains the complete slope search, subdivision, tie-breaks and generation procedure.

\paragraph{7. Rounded rows and output accumulation.}
Write $b_n(v)$ for bit $n$ of $\delta_v$, with $b_n=0$ outside $0\le n<12$. For a covered address set
\begin{align}
 d_i(v)&=-2b_{2i+1}(v)+b_{2i}(v)+b_{2i-1}(v),\quad h_i=13-2i,\quad i=0,\ldots,6,\nonumber\\
 R_s(v)&=\sum_{i=0}^{6}\operatorname{sgn}(d_i(v))
 \left\lfloor\frac{|d_i(v)|s+2^{h_i-1}}{2^{h_i}}\right\rfloor,\qquad \operatorname{sgn}(0)=0,\nonumber\\
 P(v)&=(c(v)+R_{s(v)}(v))\bmod2^{22}.
 \label{eq:delta-rows}
\end{align}
Each nonnegative magnitude is rounded to nearest with ties upward, then its sign is applied. Thus negative ties round away from zero; there is no single rounding of the complete signed product. The seven rows, including the final digit for the unsigned 12-bit offset, are added modulo $2^{22}$. Define $P(v)=C_\varnothing$ on uncovered addresses. Finally,
\begin{equation}
 \Delta(\mathbf z)=
 \begin{cases}
 C_{a_*}^{\rm call},& A\ne0,\quad a_*=\min\{a:A_a=1\},\\
 P(y+2^{20}),& A=0,\ h=1,\ y<B,\\
 C_\varnothing,&\text{otherwise},
 \end{cases}
 \qquad
 \theta_{\rm out}=(\theta_{\rm in}+\Delta(\mathbf z))\bmod2^{22}.
 \label{eq:delta-definition}
\end{equation}
No further rounding, saturation or signed-residue adjustment is applied. The definition holds for every starting accumulator; it is the additive integer action proved by Theorem~\ref{thm:factor-sum}, not the real contract payoff.

\FloatBarrier

\subsection{Autocallable: fixed-point control}
\label{app:fixed-control}
Table~\ref{tab:fixed-configurations} retains the six configurations and development outcomes. The generator made no model queries. Predeclared selection minimizes $\TQ$ among structurally valid secure-screening passes, breaking ties by qubits then identifier; it selects configuration 4.

\begin{table}[htbp]\centering\small
\caption{Fixed-point configurations and retained development results. The local table target is a fitting target in radians, not a bound on total circuit error. Counts are angle-tolerance violations among 100,000 development trajectories; screening is not a population guarantee. The selected row is bold.}\label{tab:fixed-configurations}
\begin{tabular}{@{}lrrrr@{}}\toprule
Variant & Fractional bits & Table target & Violations & Accepted $\TQ$\\\midrule
1 & 24 & $10^{-6}$ & 615 & ---\\
2 & 24 & $2.5\times10^{-6}$ & 976 & ---\\
3 & 26 & $10^{-6}$ & 0 & 12,513,540\\
\textbf{4} & \textbf{26} & $\mathbf{2.5\times10^{-6}}$ & \textbf{0} & \textbf{11,635,140}\\
5 & 28 & $10^{-6}$ & 0 & 13,044,416\\
6 & 28 & $2.5\times10^{-6}$ & 0 & 12,156,144\\\bottomrule
\end{tabular}
\end{table}

\subsection{Barrier-centred, narrow-window and signed-Booth circuits}
\label{app:other-circuits}
\paragraph{Autocallable.} The barrier-centred circuits retain $D_i(t)=F_i(t)-\tau_i(t,\ln0.7)$, the difference between an integer form and its date-dependent knock-in threshold. For the frozen hardened tape in Theorem~\ref{thm:factor-sum} (artifact path \path{optimized/autocall/hardened/r10_y0drop/candidate.py}, candidate SHA-256 \texttt{918f3703...}, tape SHA-256 \texttt{f03c6304...}), the integer map $\Delta_{\mathcal T}$ is specified explicitly as follows. The forms are $F_0(t)=K_{t,0}$, $F_1(t)=1922K_{t,0}+3329K_{t,1}$, and $F_2(t)=100336K_{t,0}+57929K_{t,1}+163848K_{t,2}$, with unit scales $k_{0,u}=0.3125L_{00}$, $k_{1,u}=0.3125L_{10}/1922$, and $k_{2,u}=0.3125L_{20}/100336$, and form sums $(n_0,n_1,n_2)=(1,5251,322113)$. Barrier and call thresholds are $T_i^B(t)=\lceil(\ln0.7-t\mu_i)/k_{i,u}+15.5\,t\,n_i\rceil$ and $T_i^C(s_a)=\lceil(\ln1.1-s_a\mu_i)/k_{i,u}+15.5\,s_a\,n_i\rceil$ for $t=1,\ldots,20$ and call dates $s_a=4(a+1)$ ($a=0,\ldots,4$). Signed widths grow per date from $(8,19,24)$ at $t=1$ to $(10,22,28)$ at $t=20$, covering all $32^{60}$ inputs without wrap; all three assets are tested for knock-in at every date via the sign bit of $D_i(t)$, and call dates compare $D_i(s_a)\ge T_i^C(s_a)-T_i^B(s_a)$. At maturity, $Y_2=\lfloor D_2(20)/2^7\rfloor$ (21-bit signed, common unit $u_C=128k_{2,u}$). $D_0(20)$ is scaled by $M_{Y_0}=6271$ (shift $S_{Y_0}=4$, column drop $G_{\mathrm{drop},0}=1$, nearest-even offset $C_{Y_0}=1190$ plus dropped-bit mean $0.5$) and truncated to 20-bit signed $Y_0$; $D_1(20)$ is scaled by $M_{Y_1}=66821$ (five-digit NAF, relative error $9.2\times10^{-7}$, shift $S_{Y_1}=18$, column drop $G_{\mathrm{drop},1}=16$, nearest-even offset $C_{Y_1}=30345$) and truncated to 20-bit signed $Y_{1,\mathrm{lo}}$, giving $y=\min(Y_0,Y_{1,\mathrm{lo}},Y_2)$. The put branch requires $A=0$, $h=1$, and $y<B_Z=20006$; the 19-bit unsigned address $U=y+2^{18}$ indexes a frozen 172-leaf minimax prefix tree (depths $\le11$, fitted on $W_{20}\ge-2.20$) stored in the artifact, returning 20-bit intercept $c_0$ and 10-bit slope $c_1$, evaluated via six radix-4 Booth rows ($K_{\max}=11$, $F=6$, nearest-integer row rounding with ties upward before sign application) and added modulo $2^{22}$ to $\theta_{\mathrm{in}}$. Uncovered addresses return $C_\varnothing=975463$, and call dates return $(C_0^{\mathrm{call}},\ldots,C_4^{\mathrm{call}})=(1057100,1144652,1243260,1365747,1647099)$. The \emph{lowest-product circuit} keeps a closely related state but uses widths that cover the loading law except in rare tails, widens registers only when needed, takes the minimum of two assets before sign extension and merges the payoff into one table stage. Its decision constants were tuned on sampled loading-law paths, which explains its higher count of selected threshold failures.

\paragraph{TARF.} The narrow-window circuits keep the cap identity of Eq.~\eqref{eq:tarf-cap-correction}. The lowest-product circuit chooses its lower index window from an absorbing-chain calculation of the probability of falling below it, decomposes the cap-date product, removes the top bit of the non-restoring square root and simplifies the lookup cleanup with CNOT basis changes. The cap-hardened circuits accumulate $T_p$ with 22 fractional bits and keep $Y$ with 18, which removes all 400 selected cap-tie failures.

\paragraph{Second setting.} The barrier-centred lattice-form circuit uses a 21-bit asset-2 virtual-register form with coefficients $(4064,2820,10505)$, defers widening of the asset-0 and asset-1 differences, and fits its per-date constants on $5\times10^7$ sampled paths.

\paragraph{HHL.} The signed-Booth circuits keep the piecewise-quadratic structure of Appendix~\ref{app:hhl} but use shifted signed-Booth coefficient words with zero-cost rounding, combine the first two Booth rows, compress the deep lookup leaves, allocate lookup outputs lazily and clear them with a decision-tree unlookup.

\begin{table}[htbp]\centering\small
\caption{Barrier-centred, narrow-window and signed-Booth circuits. $R_{\mathrm{L}}$ as in Table~\ref{tab:resources}; for HHL, $R$ is taken against the published circuit of \citet{gu}. \emph{Holdout}: angle-tolerance violations in $10^6$ task-reference paths at one recorded seed; for HHL, exhaustive over all 2,031,616 inputs. \emph{Selected}: violations on the retained stress inputs (900 threshold paths, 400 TARF cap-tie paths, 66 extreme second-setting paths). \emph{Proof}: Theorem~\ref{thm:factor-sum} or the HHL logical proof.}\label{tab:variants}
\setlength{\tabcolsep}{5pt}\renewcommand{\arraystretch}{1.05}
\begin{tabularx}{\linewidth}{@{}>{\raggedright\arraybackslash}Xrrrrrc@{}}\toprule
Circuit & $N_{\mathrm{Tof}}$ & $Q_{\max}$ & $R$ & Holdout & Selected & Proof\\\midrule
Autocallable, hardened & 6,888 & 472 & 2.87 & 0 & 6 & yes\\
Autocallable, low product & 5,109 & 442 & 4.14 & 2 & 518 & --\\
Autocallable, lowest product & 5,075 & 442 & 4.17 & 4 & 520 & --\\
\addlinespace[3pt]
TARF, cap-hardened I & 8,049 & 277 & 4.90 & 1 & 0 & --\\
TARF, cap-hardened II & 7,921 & 277 & 4.98 & 4 & 0 & --\\
TARF, low product & 7,710 & 262 & 5.40 & 1 & 54 & --\\
TARF, lowest product & 7,553 & 262 & 5.52 & 2 & 33 & --\\
\addlinespace[3pt]
Second setting, lattice forms I & 4,956 & 441 & 4.28 & 1 & 46 & --\\
Second setting, lattice forms II & 4,944 & 441 & 4.29 & 1 & 55 & --\\
\addlinespace[3pt]
HHL, signed-Booth & 965 & 116 & 1.38 & 0 & -- & --\\
HHL, signed-Booth & 972 & 115 & 1.39 & 0 & -- & yes\\
HHL, signed-Booth & 979 & 114 & 1.39 & 0 & -- & --\\
HHL, signed-Booth (minimum product) & 987 & 113 & 1.39 & 0 & -- & yes\\
HHL, signed-Booth & 999 & 112 & 1.39 & 0 & -- & --\\\bottomrule
\end{tabularx}
\end{table}

\subsection{Second-setting parameters and implementation conventions}
\label{app:second-parameters}
Table~\ref{supp:tab:generic} gives the weights for the second setting in Table~\ref{tab:contracts}. Barrier thresholds enter date-dependent offsets and sign-bit tests; call dates use shifted thresholds. Threshold fits separate distinct form values so that tied states receive the same decision.

\begin{table}[tbp]\centering\footnotesize
\caption{Second-setting integer forms $Y_i=\sum_kW_{ik}S_k$, with $S_k=\sum_{s\le t}j_s^{(k)}$. The circuits retain the weighted forms with date-dependent offsets, not separate raw-sum registers. Asset~0 has weight one on $S_0$ in both. Selected-tail outcomes are in Appendix~\ref{app:counterexamples}.}\label{supp:tab:generic}
\setlength{\tabcolsep}{5pt}
\begin{tabularx}{\linewidth}{@{}>{\raggedright\arraybackslash}p{3.2cm}>{\raggedright\arraybackslash}X>{\raggedright\arraybackslash}X@{}}\toprule
& Shift-add lattice-form circuit & Lookup lattice-form circuit\\\midrule
Asset-1 weights $(S_0,S_1)$ & 751, 2010 & 34, 91\\
Asset-2 weights $(S_0,S_1,S_2)$ & 4469, 3101, 11552 & 4616, 3203, 11932\\
Mixing at each date & Signed-digit shifted additions, packed into shared adder words & Two 32-entry lookups for factors 0 and 1; shifted additions for factor 2\\
Knock-in test & Sign bit of a register offset by the threshold & Sign bit; threshold changes carried in the lookup words\\
Put angle & Piecewise-linear table with an 8-bit address & Quadratic on 128 segments\\
\bottomrule
\end{tabularx}
\end{table}

\paragraph{Fitting and terminal conversion.}
Convolution of the loading weights assigns mass to index-sum lattice points for numerical fitting. The shift-add fit uses simultaneous approximation with a shared denominator per asset; the asset-2 denominator is near $10^4$. The lookup fit uses points within $0.003$ of the log-return threshold with marginal and joint masses above $10^{-17}$. It chooses a separating threshold when possible, otherwise minimizing misclassified mass over the retained distinct form values. These design objectives give neither an all-input guarantee nor a complete circuit-failure bound.

For shift-add conversion, let $b_{ik}=\gamma L_{ik}$ and $\lambda=W_{22}/b_{22}$. The factors $\kappa_0=\lambda b_{00}/W_{00}$ and $\kappa_1=\lambda b_{11}/W_{11}$ use 9 and 19 fractional bits, with rounded drift/centering constants and restored threshold offsets. Multiplication retains four guard bits below the common asset-2 unit before signed minimum selection; the terminal table is piecewise linear with an eight-bit address.

For lookup conversion, use regression scales $u_i=(\mathbf b_i\cdot\mathbf W_i)/(\mathbf W_i\cdot\mathbf W_i)$ on the nonzero row coefficients, and terminal offsets $c_i=20\mu_i-15.5(20)u_i\sum_kW_{ik}$. After restoring threshold offsets, convert the first two forms to the common unit $u_2$ by $u_i/u_2$. Signed-digit multipliers use $w_i+1$ fractional bits, with register widths $(w_0,w_1,w_2)=(9,15,22)$. Fixed rounded offsets compensate right-shift losses; the common word has 22 bits. The 128-segment quadratic angle table has three output guard bits and four extra curvature bits. These are finite-range, numerically fitted conventions, not a certified transcendental approximation.

\subsection{Deferred measurement-phase corrections}
\label{app:deferred-phases}
Destructive $X$-basis measurement of a lookup word $w(a,c)$, with outcome $\mathbf m$, resets its qubits and leaves $(-1)^{\mathbf m\cdot w(a,c)}$. Delayed cancellation requires the same address $a$ and controls $c$. For each fixed record of other measurements, the complete intervening block must preserve those data on the loaded-data subspace:
\[
 B=\sum_{a,c}|a,c\rangle\langle a,c|\otimes B_{a,c},\qquad
 D_{\mathbf m}=\sum_{a,c}(-1)^{\mathbf m\cdot w(a,c)}|a,c\rangle\langle a,c|,
\]
so $BD_{\mathbf m}=D_{\mathbf m}B$ for the complete block, without gate-by-gate commutation. Under a temporary permutation $\Pi$, the phase is $\Pi D_{\mathbf m}\Pi^{-1}$; restoring the original data restores the correction. Unrestored addresses or changed controls invalidate the argument.

\paragraph{HHL.} The early constant-word phase depends on the original $X$. Before final unlookup, the product block restores the coefficient word and undoes the low-input normalization, restoring that address. Saved constant-word outcomes combine with the remaining word measurements. The complete block preserves $X$; Appendix~\ref{sec:hhlproof} gives the structural check.

\paragraph{TARF.} Reverse dates undo later index updates, restoring $K_r=K_t+\mathrm{OFF}_t\pmod{2^7}$. At dates 2--25, saved stop-date data reconstruct the stop bit and alive control; post-date accrual words reconstruct the cap decision. Date 26 leaves stop/alive data unchanged. After restoring these controls, the reverse step reloads the same word, subtracts its contributions and combines forward/reverse outcomes in one phase lookup before undoing the index increment. The first-date inverse restores stop/alive data and unlooks up its word before clearing the index. These conditions justify local correction, not an all-input structural certificate for TARF.

\subsection{Reference and final-circuit stage accounting}
\label{app:stage-accounting}
\begin{figure}[htbp]\centering\begin{tikzpicture}
\begin{axis}[paperaxis,xbar stacked,width=0.74\linewidth,height=5.4cm,bar width=11pt,
xmin=0,xmax=33000,ymin=-0.6,ymax=3.6,ytick={0,1,2,3},
yticklabels={{TARF: windowed circuit},{TARF: lookup reference},{Autocall: factor-sum circuit},{Autocall: log-return reference}},
xlabel={Toffoli gates},xmajorgrids=true,legend style={font=\footnotesize,draw=none,at={(0.5,1.03)},anchor=south},legend columns=3,
scaled x ticks=false,xticklabel style={/pgf/number format/1000 sep={\,}}]
\addplot[fill=Sea!75,draw=Sea] coordinates {(3097,0) (11268,1) (3088,2) (7585,3)};
\addplot[fill=Gold!65,draw=Gold] coordinates {(1733,0) (5892,1) (1375,2) (1650,3)};
\addplot[fill=Sea!30,draw=Sea] coordinates {(3551,0) (11268,1) (3088,2) (7575,3)};
\legend{Forward dates,Payoff stage,Reverse dates}
\node[font=\footnotesize,anchor=west] at (axis cs:8381,0) {8,381};
\node[font=\footnotesize,anchor=west] at (axis cs:28428,1) {28,428};
\node[font=\footnotesize,anchor=west] at (axis cs:7551,2) {7,551};
\node[font=\footnotesize,anchor=west] at (axis cs:16810,3) {16,810};
\end{axis}\end{tikzpicture}
\caption{Toffoli accounting by forward dates, terminal payoff and reverse dates for each reference/final pair. Bars sum to the retained secure totals. These are implementation comparisons, not controlled ablations; Table~\ref{tab:stage-accounting} gives the combined detailed stage counts.}
\label{fig:profile}\end{figure}
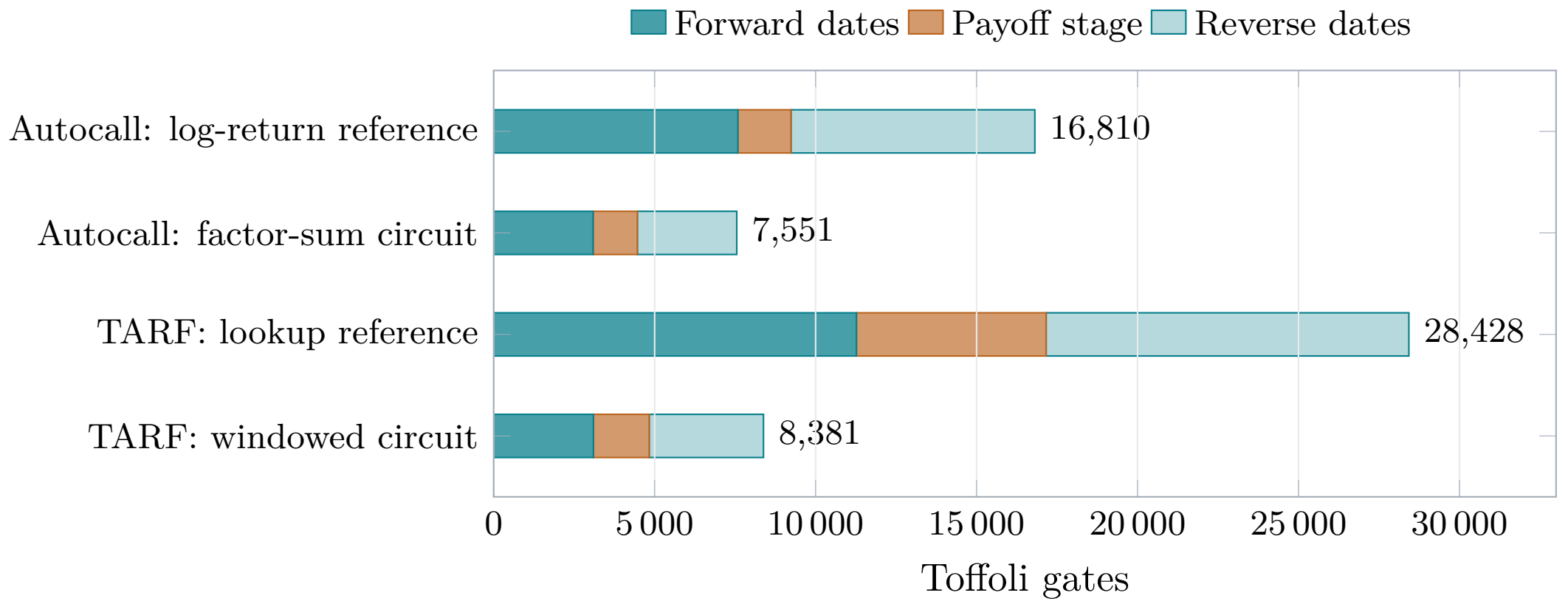

Table~\ref{tab:stage-accounting} gives the saved profiles behind Figure~\ref{fig:profile}. Instrumented builds match the retained program identities and secure totals.

\begin{table}[htbp]\centering\small
\caption{Retained Toffoli stage accounting: (a) autocallable and (b) TARF, each comparing its reference and final circuit. Date stages sum both sweeps; each column sums to its secure total. These are implementation comparisons, not controlled ablations.}\label{tab:stage-accounting}\label{tab:autostages}\label{tab:tarfstages}
\setlength{\tabcolsep}{3pt}
\begin{minipage}[t]{0.49\linewidth}\centering
\textbf{(a) Autocallable}\par\smallskip
\begin{tabularx}{\linewidth}{@{}>{\raggedright\arraybackslash}Xrr@{}}\toprule
Stage & \shortstack{Log-return\\reference} & \shortstack{Factor-sum\\circuit}\\\midrule
Path: table lookups & 3,600 & ---\\
Path: lookup clean-up & 3,600 & ---\\
Path: register adds & 6,960 & 894\\
Asset forms & --- & 4,398\\
Barrier and call tests & 1,000 & 884\\
Payoff: worst-of forms & --- & 460\\
Payoff: other & 1,650 & 915\\\midrule
Toffoli total & 16,810 & 7,551\\\bottomrule
\end{tabularx}
\end{minipage}\hfill
\begin{minipage}[t]{0.49\linewidth}\centering
\textbf{(b) TARF}\par\smallskip
\begin{tabularx}{\linewidth}{@{}>{\raggedright\arraybackslash}Xrr@{}}\toprule
Stage & \shortstack{Lookup\\reference} & \shortstack{Windowed\\circuit}\\\midrule
Index-sum update & 676 & 300\\
Payment lookup & 4,562 & 3,738\\
Lookup clean-up & 4,562 & 453\\
Accumulator adds & 7,384 & 2,072\\
Knock-out, cap and stop logic & 5,352 & 85\\
Cap-date correction & 2,058 & 778\\
Square root & --- & 414\\
Angle table & 3,834 & 541\\\midrule
Toffoli total & 28,428 & 8,381\\\bottomrule
\end{tabularx}
\end{minipage}
\end{table}
\FloatBarrier
\label{budget:A-integer-arithmetic:end}
\section{Proofs and validation methods}
\label{app:validation}
\subsection{Statistical methods and simulator validation}
\label{app:statistics}
For a circuit fixed before independent validation draws, use the zero-failure bound in Section~\ref{sec:independent-validation}. The finance studies sample $p_{\mathrm{load}}$, uniform 22-bit incoming accumulators and evaluator measurement outcomes, unless the all-input theorem removes a dependence. The output event is an angle-tolerance violation; bit-for-bit simulator disagreement and sampled structural failures are separate events.

Table~\ref{tab:multiplicity} applies that bound at the displayed per-circuit $\alpha$ within each declared comparison family, giving simultaneous one-sided 95\% coverage. Separate references, studies, development samples and selected inputs are not pooled. Reusing the three fixed second-setting samples does not create new draws.

Independent-simulator replays for the factor-sum and windowed TARF circuits give bit-for-bit agreement on one million inputs each, using the same task model; they cross-check simulation, not the financial reference.

\paragraph{Barrier-centred, narrow-window and signed-Booth circuits.} These circuits were designed with access to the retained circuits, proofs and selected stress inputs, so their selected-input counts are not held out. Each finance circuit has one million-path task-reference holdout at a seed recorded before the run but not fixed in version control; it is a fresh seed, not a verifiable preregistration. For $k=0,1,2,4$ violations in $10^6$ paths, the one-sided 95\% Clopper--Pearson upper rates are $3.00$, $4.74$, $6.30$ and $9.15\times10^{-6}$; these per-circuit bounds are not multiplicity-adjusted, and no independent 60/90-digit study is available. Their secure counts were reproduced on two further seeds, and the unchanged checker reproduced their selected-input counts; on the same inputs, the search circuits return the counts of Appendix~\ref{app:counterexamples}.

\begin{table}[htbp]\centering\small
\caption{Declared zero-in-$10^6$ Bonferroni bounds, applied separately to each reference/study. Each family has simultaneous one-sided 95\% coverage. Failed or unfilled predeclared search slots remain included. Individual TARF statements instead use $\alpha=0.05$ for one preselected circuit and event; no joint statement across TARF circuits or all studies is asserted.}\label{tab:multiplicity}
\begin{tabularx}{\linewidth}{@{}>{\raggedright\arraybackslash}Xrr@{}}\toprule
Comparison family & Per-circuit $\alpha$ & Upper rate\\\midrule
Original factor-sum and log-return pair & $0.05/2$ & $3.68887\times10^{-6}$\\
Pair plus fixed-point control; separately, three fixed transfers & $0.05/3$ & $4.09434\times10^{-6}$\\
Fresh second-setting search (four slots) & $0.05/4$ & $4.38202\times10^{-6}$\\
Three fixed plus four fresh slots & $0.05/7$ & $\GenJoint$\\\bottomrule
\end{tabularx}
\end{table}
The Qualtran kickmix conversion~\citep{khattar2026} reproduces secure counts and gives zero sampled phase, input and workspace errors in its separate million-input studies. Five deliberately injected faults per circuit (missing phase fixes, a dirty reset, a dirty final qubit, a flipped input and a flipped output) were all detected. Sampled structural success does not close all-input obligations.

\subsection{Selected autocallable counterexamples}
\label{app:counterexamples}
The selected stress population has 900 paths near call, barrier and strike thresholds; the 901-fixture replay adds the distinct historical search failure. Among the 900 deliberately selected paths, the factor-sum circuit has 241 angle-tolerance violations: 79 call-decision cases, 116 knock-in-decision cases, 8 numerical-angle cases and 38 terminal-wrap cases; these diagnostic counts are not loading-law failure probabilities. Table~\ref{tab:selected-auto} records the other selected-fixture outcomes. Approximate integer coefficients cause decision errors. Signed terminal overflow can create a false worst asset or select the wrong table leaf. The second-setting fitted weights sacrifice rare tail inputs for lower arithmetic cost.
\begin{table}[htbp]\centering\small
\caption{Other selected autocallable fixture outcomes: angle-tolerance violations. The original factor-sum archive separately retains one allowed violation in 100,000 screening paths. No archived circuit was repaired.}\label{tab:selected-auto}
\begin{tabularx}{\linewidth}{@{}>{\raggedright\arraybackslash}X>{\raggedright\arraybackslash}p{3.9cm}r@{}}\toprule
Circuit / diagnosis & Selected population & Violations\\\midrule
Log-return reference; fixed-point control & Same 900 paths, separately & 52 each\\
All second-setting designs & Second-setting 900 boundary paths & 0 each\\
Shift-add / lookup lattice forms & 66 extreme second-setting inputs & 49 / 46\\
Transferred log-return; other two fixed controls & Same 66 inputs, separately & 26; 0 / 0\\\bottomrule
\end{tabularx}
\end{table}

\subsection{Factor-sum proof support}
\label{app:proof-records}
For the frozen factor-sum tape identified in the artifact manifest, the hooked and separately scored tapes must coincide. Cuts isolate each date's sum update and decision block, terminal materialization $S_1$, branch logic $S_2$ and eight accumulator adders. Assertions check the live registers at adder boundaries and the complement gates implementing signed addition via $\lnot(\lnot x+v)=x-v\bmod2^{22}$. Every live slot initially denotes an independent variable on its declared domain; internal allocations start at zero.

Table~\ref{tab:outputproof} gives the output-value obligations. Enumeration exhausts the stated finite domains; cancellation is exact canonical XOR--AND cancellation against a reference adder for all operand values. Each lemma checks that outputs depend only on declared inputs, other live registers are unchanged and internal frees are zero. The terminal market-state and history-flag checks separate by verified support. Structural obligations additionally establish phase cancellation and persistent-state restoration; these do not determine output values. All 92 value lemmas close without SAT; structural phase/restoration checks also use SAT refutations accepted by the independent DRAT checker.

Write $\kappa(t)=t+\max(0,t-11)$ for the breach-counter bound from Appendix~\ref{app:integer-model}, and $k_q=\mathbf1\{K_C=0\}$ for the terminal no-knock bit. The call flags are $A_a=\mathbf1\{C_a=3\}$.

\begin{table}[htbp]\centering\small
\caption{Factor-sum output-value lemma families (92 lemmas, all proved). Domains are exhaustive; terminal component domains combine through the verified support separation. Execution totals are retained in the certificate inventory.}\label{tab:outputproof}
\begin{tabularx}{\linewidth}{@{}>{\raggedright\arraybackslash}Xr>{\raggedright\arraybackslash}p{2.0cm}@{}}\toprule
Statement and domain & Lemmas & Proof method\\\midrule
Layout: allocation-only setup; only the eight adder ranges write $\theta$ & 1 & Syntactic\\
Date/factor updates, $t=1,\ldots,20$, $k=0,1,2$: $K_{t-1,k}\in[0,31(t-1)]$, $j_t^{(k)}\in[0,31]$ give $K_{t,k}\in[0,31t]$ & 60 & Enumeration\\
Date decisions: $\mathbf K_t\in[0,31t]^3$, $K_C\in[0,\kappa(t-1)]$ give $K_C'\in[0,\kappa(t)]$ and the specified call count. At a call date its counter starts at zero; earlier counters lie in $[0,3]$, later counters remain zero & 20 & Enumeration\\
Terminal $S_1$: $\mathbf K\in[0,620]^3$; separately $K_C\in[0,29]$, $C_a\in[0,3]$. Compute signed $y$, $[y<B]$, $k_q$ and the five $A_a$ flags & 2 & Enumeration\\
Each accumulator adder: $\theta\mathrel{+}=\pm(a+c_{\mathrm{in}})\bmod2^{22}$ for every operand and incoming accumulator, preserving the operands & 8 & Cancellation\\
Terminal $S_2$: the eight addends sum to the model value for every $y\in[-2^{20},2^{20})$, with predicate $[y<B]$, and every $k_q\in\{0,1\}$ and $(A_0,\ldots,A_4)\in\{0,1\}^5$ (64 flag combinations) & 1 & Enumeration\\\bottomrule
\end{tabularx}
\end{table}

Initialization supplies zero sums/counters; each date update supplies the decision block's domain, whose outputs satisfy the next date's domain. The terminal outputs supply $S_2$'s full signed-word and flag domain, so no reachable-image fallback is used. The all-operand adder identities compose with the enumerated addend sum to give
\begin{equation}
 \theta_{\mathrm{out}}=\theta+\Delta(\mathbf z)\bmod2^{22}.
 \label{eq:outputproof}
\end{equation}
Restoration composes outward from the terminal block: it preserves every live register except $\theta$ for arbitrary cut-point values, and paired forward/reverse dates restore their live registers and clear temporary work. The initially zero workspace therefore returns to zero. Phase cancellation under Section~\ref{sec:outputproof}'s measurement semantics supplies the separate coherent-action argument.

The trusted computational base comprises the mirrored gate/measurement semantics, frozen integer model/data, asserted segmentation, canonical symbolic and bitsliced evaluators, Python/NumPy, Tseitin encoding and independent DRAT checker. This is a computational certificate, not a proof-assistant formalization.

\paragraph{Restoration and phases.} Restoration composes from the terminal block outwards: the block preserves every live register except $\theta$ for arbitrary cut-point values, and each forward date paired with its reverse restores its live registers and clears its temporaries. Phase obligations close by canonical cancellation or by unsatisfiability certificates accepted by the independent DRAT checker. Only diagonal gates are measurement-conditioned, so basis values are record-independent and the sign splits into an unconditioned Boolean function plus one coefficient per measurement bit; all vanish. For a measured basis bit $q$ and outcome $m$, \texttt{hmr} contributes $2^{-1/2}(-1)^{mq}$, which gives the branch identity $K_{\mathbf m}=2^{-M/2}U_\Delta$ of Theorem~\ref{thm:factor-sum}. Basis-value agreement alone would not establish this conclusion.

\paragraph{Hardened circuit.} The hardened tape is cut at the same date boundaries. A syntactic pass shows that all 300 inputs are preserved, the unconditioned phase vanishes and all 6,679 frees are syntactically zero; the 31 remaining measurement-phase coefficients vanish on all 4,096 points of their support. Eight accumulator ranges cancel exactly against a reference adder. Date-update lemmas combine exhaustive evaluation with worst-case range bounds, and the terminal lemmas enumerate all reachable control states; no SAT refutation is needed. The proof scripts were adapted from the factor-sum and HHL proofs and rerun independently with identical results, but no proof protocol was fixed before the runs and the composition argument has not had an external review.

\paragraph{Reference-circuit scope.}
For the log-return and TARF lookup references, inputs are preserved and the unconditioned phase vanishes, but restoration of long-lived state after the reverse sweep remains open. These obligations require reachable-state invariants and wide addition identities; an open obligation is not a demonstrated failure. A TARF per-date decomposition with the accrual-counter invariant closes its counter conditions, but partial closure is not a complete certificate. The artifact manifest identifies the full residual inventory and certificate records.

\subsection{Implementation checks and numerical records}
\subsubsection{Independent-reference TARF holdout records}
\label{app:tarf-truth}
The windowed TARF candidate was fixed before the study. An independent reference was typed from the contract specification and imports neither the task model nor a candidate. The study sampled $10^6$ independent paths from $p_{\mathrm{load}}$, uniform 22-bit incoming accumulators and evaluator measurement outcomes, recovering unsigned increments modulo $2^{22}$. Every path was evaluated at both 60 and 90 decimal digits, with ten guard digits.

\begin{table}[htbp]\centering\small
\caption{Independent-reference windowed TARF holdout outcomes.}\label{tab:tarf-truth-record}
\begin{tabularx}{\linewidth}{@{}>{\raggedright\arraybackslash}p{3.8cm}>{\raggedright\arraybackslash}X@{}}\toprule
Quantity & Observed result\\\midrule
Angle-tolerance violations & $0/10^6$\\
Maximum angle error & $7.93505\times10^{-6}$\,rad\\
Sampled structural errors & Zero input modifications, residual phases, dirty frees and nonzero final workspace\\
One-sided 95\% upper rate & $\TarfTruthUpper$\\\bottomrule
\end{tabularx}
\end{table}
The Clopper--Pearson bound is for one preselected circuit and primary event, conditional on the numerical reference, without pooling or multiplicity adjustment. Precision agreement is not a certified reference enclosure; sampled structural success is not an all-input proof.

\subsubsection{Logical-inverse implementation and replay}
\label{app:inverse-check}
The wrappers add two accumulator-complement layers while preserving the forward body and ordinary measurement corrections. Bounded replays checked the negative increment and both composition orders, $U^{-1}U$ and $UU^{-1}$, on ordinary and boundary inputs with varied incoming accumulators and independently sampled measurement records. No errors occurred in increments, either composition, input preservation, phase cancellation, clean frees or final workspace. The all-input inverse on clean workspace follows from Corollary~\ref{cor:inverse} and the additive-action proofs; replay only checks implementation. Controlled use, arbitrary initialized workspace and full pricing remain outside these checks.

\begin{table}[htbp]\centering\small
\caption{Retained inverse-wrapper resources; only accumulator-complement $X$ gates are added.}\label{tab:inverse-check}
\begin{tabular}{@{}lrrr@{}}\toprule
Circuit & $N_{\mathrm{Tof}}$ & $Q_{\max}$ & Added $X$\\\midrule
Factor-sum & 7,551 & 518 & 44\\
HHL & 1,057 & 126 & 66\\\bottomrule
\end{tabular}
\end{table}
\FloatBarrier

\subsubsection{Lattice calculation: numerical reliability}
\label{app:lattice-records}
The recursion compares the proved integer model with the financial target. Each factor axis is cropped to retain all but $10^{-16}$ of its marginal law; discarded mass is accumulated across dates. At a one-sided call, the angle word is separated from every remaining outcome of the other side by more than $7\times10^4$ units, versus tolerance $10.48576$ units, justifying absorption as a violation. The omission allowance does not enclose numerical error.

The vectorized output agrees bit for bit with the integer emulator on all 901 stress/historical fixtures and 20,000 random paths; the emulator agrees with gate-level outputs throughout the million-path study and on all fixtures. The minimum decision margin is approximately $4.3\times10^{-8}$, and the smallest gap between an output deviation and the tolerance exceeds $2\times10^{-3}$ angle-word units. These margins support the numerical classifications but are not an outward-rounded certificate. Independent Monte Carlo agrees with the decision-only result; a separate $5\times10^8$-path study supports the full-output probability and bias.

Probability propagation uses nonnegative terms, about $2\times10^3$ operations per point followed by sums of at most $1.25\times10^7$ terms. With unit roundoff $2^{-53}$, the a priori relative-roundoff estimate is far below $10^{-8}$. Signed bias sums also evaluate transcendental functions. The probability allowance covers omitted mass; the bias allowance covers omitted and call-disagreement mass. Neither certifies all roundoff, transcendental or classification error.

The same recursion applies to the hardened integer model, which tests every asset at every date. Its calculation gives a decision-error probability of zero up to omitted mass $4.7\times10^{-13}$ and a full violation probability of $3.5555\times10^{-8}$, all from paths with $W_{20}<-2.20$ in the far put tail. On correct lattice paths its largest deviation is 10.4853 units, only $5\times10^{-4}$ units inside the tolerance.

\begin{figure}[!htb]\centering\begin{tikzpicture}
\begin{semilogxaxis}[paperaxis,width=0.64\linewidth,height=5.8cm,xmin=1e-8,xmax=3e-5,ymin=-0.6,ymax=5.6,
ytick={0,1,2,3,4,5},yticklabels={{Lattice value, full output},{Lattice value, decisions only},{Monte Carlo $5{\times}10^8$, full output},{Monte Carlo $10^9$, decisions only},{Sampled bound, $10^6$ error-free paths},{Task budget}},
xlabel={Angle-tolerance-violation probability},xmajorgrids=true,title={(a) Lattice values, Monte Carlo estimates and upper bounds},title style={font=\small\bfseries}]
\addplot[only marks,mark=*,mark size=3pt,Sea] coordinates {(6.247569905725987e-08,0)};
\node[font=\footnotesize,anchor=west,text=Sea,fill=white,inner sep=1pt] at (axis cs:7.809462382157484e-08,0) {$6.25\times10^{-8}$};
\addplot[only marks,mark=*,mark size=3pt,Sea] coordinates {(6.032740017594563e-08,1)};
\node[font=\footnotesize,anchor=west,text=Sea,fill=white,inner sep=1pt] at (axis cs:7.540925021993204e-08,1) {$6.03\times10^{-8}$};
\draw[Ink,thick,{|[width=5pt]}-{|[width=5pt]}] (axis cs:3.4764251683501746e-08,2) -- (axis cs:6.983216033984813e-08,2);
\addplot[only marks,mark=o,mark size=2.6pt,Ink,thick,fill=white] coordinates {(5e-08,2)};
\draw[Ink,thick,{|[width=5pt]}-{|[width=5pt]}] (axis cs:4.964131580299665e-08,3) -- (axis cs:7.659895137197811e-08,3);
\addplot[only marks,mark=o,mark size=2.6pt,Ink,thick,fill=white] coordinates {(6.2e-08,3)};
\draw[Ink!70,very thick] (axis cs:2.995727786352544e-06,3.7199999999999998) -- (axis cs:2.995727786352544e-06,4.28);
\draw[Ink!70,thick,-{Stealth[length=2mm]}] (axis cs:2.995727786352544e-06,4) -- (axis cs:4.992879643920907e-07,4) node[left,font=\footnotesize,text=Ink!70] {upper bound $3.0\times10^{-6}$};
\draw[Gold,very thick] (axis cs:1e-05,4.72) -- (axis cs:1e-05,5.28);
\draw[Gold,thick,-{Stealth[length=2mm]}] (axis cs:1e-05,5) -- (axis cs:1.6666666666666669e-06,5) node[left,font=\footnotesize,text=Gold] {requirement $1.0\times10^{-5}$};
\addplot[Gold,dashed] coordinates {(1e-5,-0.6) (1e-5,5.6)};
\end{semilogxaxis}\end{tikzpicture}
\par\medskip
\begin{tikzpicture}
\begin{semilogxaxis}[paperaxis,width=0.64\linewidth,height=4.4cm,xmin=1e-13,xmax=1e-5,ymin=-0.6,ymax=3.6,
ytick={0,1,2,3},yticklabels={{Call mismatch},{Value or wrap},{False knock-in},{Missed knock-in}},
xlabel={Probability under the loading law},xmajorgrids=true,title={(b) Lattice values by mechanism},title style={font=\small\bfseries}]
\addplot[only marks,mark=*,mark size=3pt,Gold] coordinates {(3.170995611527946e-12,0)};
\node[font=\footnotesize,anchor=west,text=Gold,fill=white,inner sep=1pt] at (axis cs:5.073592978444714e-12,0) {$3.2\times10^{-12}$};
\addplot[only marks,mark=*,mark size=3pt,Gold] coordinates {(2.1482989842837973e-09,1)};
\node[font=\footnotesize,anchor=west,text=Gold,fill=white,inner sep=1pt] at (axis cs:3.4372783748540757e-09,1) {$2.1\times10^{-9}$};
\addplot[only marks,mark=*,mark size=3pt,Gold] coordinates {(2.3374143858639536e-08,2)};
\node[font=\footnotesize,anchor=west,text=Gold,fill=white,inner sep=1pt] at (axis cs:3.739863017382326e-08,2) {$2.3\times10^{-8}$};
\addplot[only marks,mark=*,mark size=3pt,Gold] coordinates {(3.6950085218725004e-08,3)};
\node[font=\footnotesize,anchor=west,text=Gold,fill=white,inner sep=1pt] at (axis cs:5.9120136349960006e-08,3) {$3.7\times10^{-8}$};
\end{semilogxaxis}\end{tikzpicture}
\caption{Factor-sum angle-tolerance violations under $p_{\mathrm{load}}$. (a) Filled points: numerical lattice values, with omitted mass bounded by $\ExactLost$ but no certified numerical enclosure. Hollow points/bars: independent full-output ($5\times10^8$) and decision-only ($10^9$) Monte Carlo estimates with 95\% intervals; arrows mark the single-circuit 95\% zero-in-$10^6$ bound and task budget. (b) Lattice values by mechanism. Numerical records are in Appendix~\ref{app:lattice-records}.}
\label{fig:exact}\end{figure}
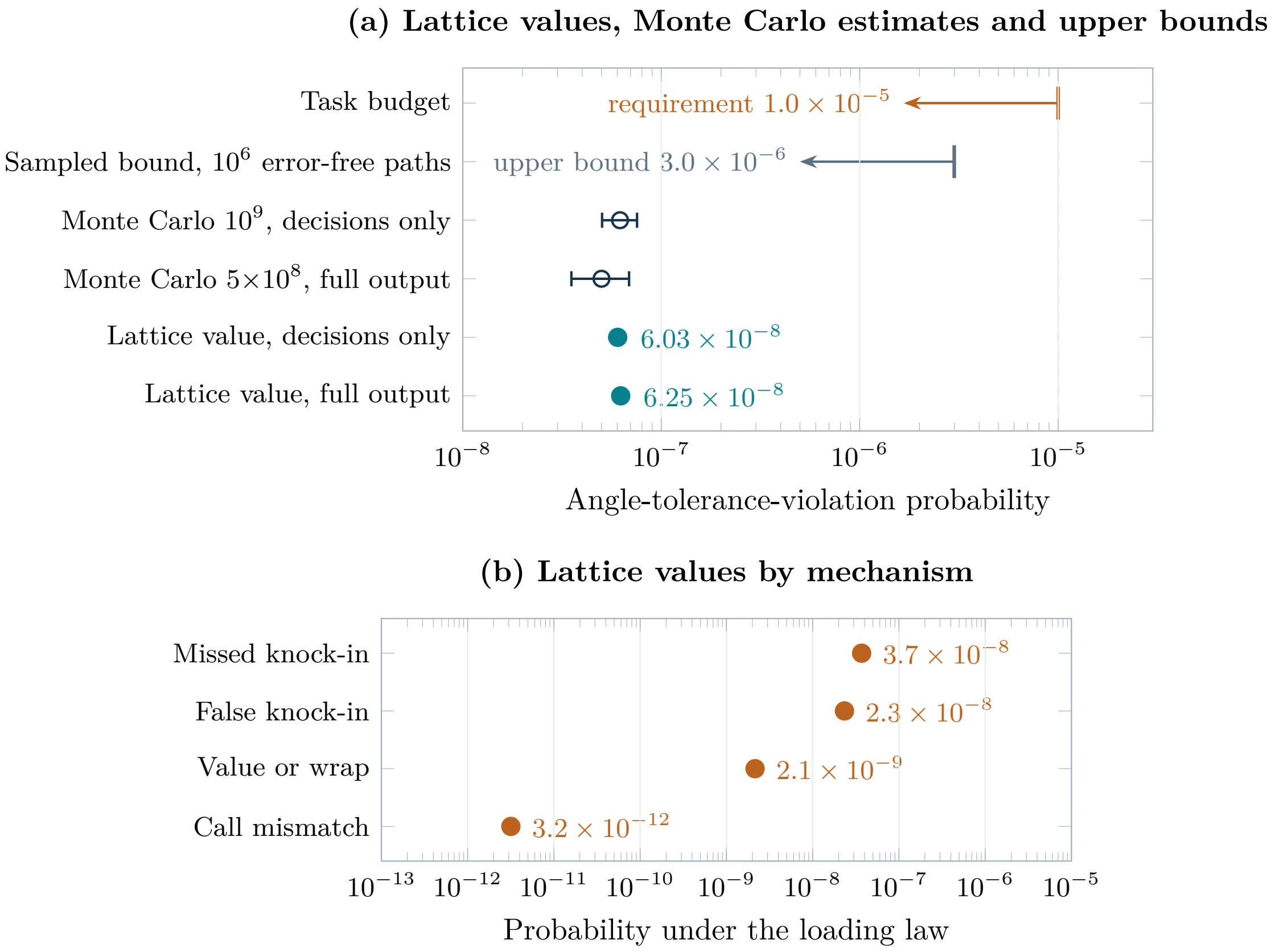

\subsection{HHL calibration}
\label{app:hhl}
\subsubsection{Task, construction, and resources}
\label{app:extra}
The calibration task of \citet{gu} preserves a 21-bit input $2^{16}\le X<2^{21}$ (2,031,616 values) and adds a 33-bit angle word to an arbitrary accumulator $k$. The target is $a(X)=\arcsin(2^{15}/X)$, scaled by $2^{32}$, with tolerance $10^{-7}$\,rad. The Booth-polynomial circuit came from an assisted search (Appendix~\ref{app:search-protocols}).

Piecewise-quadratic evaluation~\citep{hrs} partitions the domain into 222 aligned leaves of $2^L$ inputs, $9\le L\le15$. A lookup supplies constant, linear and curvature words and shift $15-L$. The constant is added and measured; controlled swaps normalize the low input window, curvature arithmetic updates the linear word, and seven radix-4 Booth rows~\citep{booth,macsorley} add the product. Normalization and curvature updates are reversed, and measurement-based unlookup~\citep{gidney,berry2019} clears the coefficients. Appendix~\ref{app:deferred-phases} gives the deferred-phase conditions. The frozen integer model $f_{\mathrm{int}}(X)$ includes row floors, finite windows and carry/borrow corrections; fitting accounts for this arithmetic without certifying transcendental error. The artifact manifest identifies the frozen coefficient data.

\begin{figure}[htbp]\centering
\begin{tikzpicture}
\begin{semilogxaxis}[paperaxis,axis on top=false,width=0.94\linewidth,height=5.8cm,log ticks with fixed point,xticklabel style={/pgf/number format/1000 sep={\,}},title style={font=\small\bfseries},xlabel={Toffoli gates $N_{\mathrm{Tof}}$},ylabel={Peak live qubits $Q_{\max}$},grid=major,major grid style={Ink!8},clip mode=individual,height=5.2cm,title={(a) HHL angle (calibration)},
xmin=200,xmax=30000,ymin=80,ymax=350,xtick={300,1000,3000,10000,30000}]
\addplot[Ink!30,dashed,thin,smooth,forget plot] coordinates {(442.9,350.00) (492.2,314.98) (546.9,283.47) (607.7,255.10) (675.2,229.58) (750.3,206.61) (833.7,185.94) (926.4,167.33) (1029.4,150.59) (1143.9,135.52) (1271.0,121.96) (1412.4,109.76) (1569.4,98.78) (1743.9,88.89) (1937.7,80.00)};
\addplot[Ink!30,dashed,thin,smooth,forget plot] coordinates {(318.7,350.00) (354.1,314.98) (393.5,283.47) (437.2,255.10) (485.8,229.58) (539.8,206.61) (599.8,185.94) (666.5,167.33) (740.6,150.59) (823.0,135.52) (914.5,121.96) (1016.1,109.76) (1129.1,98.78) (1254.6,88.89) (1394.1,80.00)};
\addplot[only marks,mark=square*,mark size=2.6pt,Ink] coordinates {(14922,323) (2163,193)};
\addplot[only marks,mark=*,mark size=2.1pt,Sea,fill=white,thick] coordinates {(1482,151) (18837,97) (2537,178) (2903,152)};
\addplot[only marks,mark=diamond*,mark size=3.4pt,Gold] coordinates {(1348,115)};
\addplot[only marks,mark=*,mark size=3.0pt,Sea] coordinates {(1057,126)};
\addplot[only marks,mark=star,mark size=3.6pt,Purple,very thick] coordinates {(987,113)};
\draw[Ink!70,thin] (axis cs:925,109) rectangle (axis cs:1380,129);
\node[font=\footnotesize\bfseries,text=Ink,anchor=south,inner sep=2pt] at (axis cs:1140,129) {(b)};
\node[font=\footnotesize,fill=white,fill opacity=0.9,text opacity=1,inner sep=1.5pt,anchor=north east,text=Ink,xshift=-2pt] at (axis cs:14922,323) {HRS reconstruction};
\node[font=\footnotesize,fill=white,fill opacity=0.9,text opacity=1,inner sep=1.5pt,anchor=south east,text=Ink,xshift=-2pt] at (axis cs:2163,193) {Dyadic reference};
\end{semilogxaxis}
\end{tikzpicture}
\par\medskip
\begin{tikzpicture}
\begin{axis}[paperaxis,axis on top=false,width=0.94\linewidth,height=5.8cm,xticklabel style={/pgf/number format/1000 sep={\,}},title style={font=\small\bfseries},xlabel={Toffoli gates $N_{\mathrm{Tof}}$},ylabel={Peak live qubits $Q_{\max}$},grid=major,major grid style={Ink!8},clip mode=individual,height=5.0cm,title={(b) Detail of the boxed region},
xmin=925,xmax=1380,ymin=109,ymax=129,xtick={950,1000,1050,1100,1150,1200,1250,1300,1350},ytick={110,115,120,125}]
\addplot[Ink!30,dashed,thin,smooth,forget plot] coordinates {(1201.7,129.00) (1214.4,127.65) (1227.2,126.32) (1239.9,125.03) (1252.6,123.75) (1265.4,122.51) (1278.1,121.29) (1290.9,120.09) (1303.6,118.92) (1316.3,117.77) (1329.1,116.64) (1341.8,115.53) (1354.5,114.45) (1367.3,113.38) (1380.0,112.33)};
\addplot[Ink!30,dashed,thin,smooth,forget plot] coordinates {(925.0,120.57) (932.0,119.67) (939.0,118.77) (946.0,117.89) (953.1,117.02) (960.1,116.17) (967.1,115.33) (974.1,114.50) (981.1,113.68) (988.1,112.87) (995.2,112.07) (1002.2,111.29) (1009.2,110.52) (1016.2,109.75) (1023.2,109.00)};
\addplot[only marks,mark=diamond*,mark size=3.4pt,Gold] coordinates {(1348,115)};
\addplot[only marks,mark=*,mark size=3.0pt,Sea] coordinates {(1057,126)};
\addplot[only marks,mark=star,mark size=2.6pt,Purple,very thick] coordinates {(965,116) (972,115) (979,114) (987,113) (999,112)};
\addplot[only marks,mark=o,mark size=4.3pt,Purple,thin] coordinates {(972,115) (987,113)};
\node[font=\footnotesize,fill=white,fill opacity=0.9,text opacity=1,inner sep=1.5pt,anchor=east,text=Gold!85!black,xshift=-6pt] at (axis cs:1348,115) {Published circuit~\citep{gu}};
\node[font=\footnotesize,fill=white,fill opacity=0.9,text opacity=1,inner sep=1.5pt,anchor=north west,text=Sea,xshift=4pt,yshift=-3pt] at (axis cs:1057,126) {Booth-polynomial circuit};
\node[font=\footnotesize,fill=white,fill opacity=0.9,text opacity=1,inner sep=1.5pt,anchor=west,text=Purple,xshift=8pt] at (axis cs:999,112) {Signed-Booth circuits};
\node[font=\scriptsize,text=Ink!60,anchor=west,inner sep=1pt] at (axis cs:1212,127.6) {equal $\TQ$ to \citet{gu}};
\node[font=\scriptsize,text=Ink!60,anchor=south west,inner sep=1pt] at (axis cs:932,120.3) {equal $\TQ$ to $987\times113$};
\end{axis}
\end{tikzpicture}
\caption{HHL angle-calibration resources. (a) Full plane; (b) enlargement of the boxed region. The filled teal circle is the Booth-polynomial circuit, which has exact all-input structural/output proofs and an exhaustive 45-digit-reference accuracy comparison (Section~\ref{sec:hhlproof}); the latter is numerical, without a certified enclosure. Hollow teal points have empirical screening evidence only. Squares identify the HRS-based reconstruction and interactive dyadic reference; the orange diamond is the best published circuit of \citet{gu}. Stars are signed-Booth circuits (Appendix~\ref{app:other-circuits}); all pass the exhaustive output check, and the ringed $987\times113$ and $972\times115$ circuits also have proofs. Every signed-Booth circuit uses fewer Toffoli gates than the published circuit, and four of the five use no more qubits (the $965\times116$ variant uses one more qubit). Dashed curves have equal $\TQ$; position ranks resources, not accuracy.}
\label{fig:resources-extra}\end{figure}
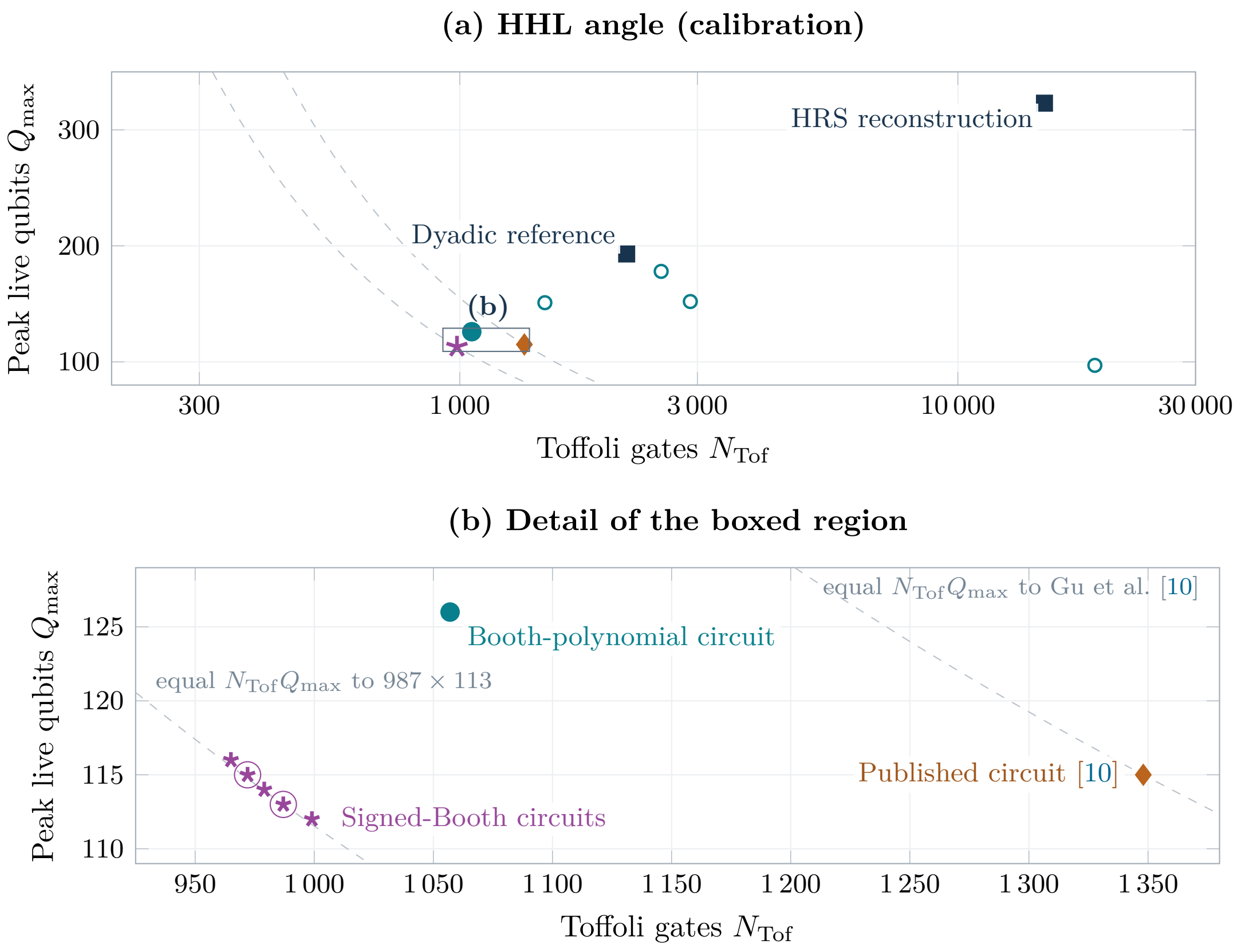

Table~\ref{tab:hhl-resources} and Figure~\ref{fig:resources-extra} compare this angle component. The H\"aner--Roetteler--Svore (HRS) reconstruction~\citep{hrs} has a product close to the published $4.8\times10^6$; the Booth-polynomial circuit is 14.1\% below the best published product, with eleven more qubits. The signed-Booth circuits of Appendix~\ref{app:other-circuits} remove this tradeoff: the minimum-product circuit is 28\% below the published product with two fewer qubits. These counts do not establish search reliability or a complete HHL implementation.
\begin{table}[htbp]\centering\small
\caption{HHL angle-task resource comparison. Reconstruction and final-circuit counts are retained secure counts; the published point is from \citet{gu}.}\label{tab:hhl-resources}
\begin{tabular}{lrrr}\toprule
Circuit & Toffoli gates & Peak qubits & $\TQ$\\\midrule
HRS-based reconstruction & 14,922 & 323 & 4,819,806\\
Best published point & 1,348 & 115 & 155,020\\
Booth-polynomial circuit & 1,057 & 126 & 133,182\\
Signed-Booth circuit (minimum product) & 987 & 113 & 111,531\\\bottomrule
\end{tabular}
\end{table}

\FloatBarrier

\subsubsection{Logical verification and numerical accuracy}
\label{sec:hhlproof}
The twelve exact logical lemmas in Table~\ref{supp:tab:hhlproof} establish phase cancellation, clean workspace and the integer increment. Phase coefficients cancel symbolically or vanish on their complete address support; all-operand adder identities and the exhaustive addend sum establish the output. Under Section~\ref{sec:outputproof}'s measurement semantics, for every allowed $X$, every accumulator $k$ and every measurement record,
\[
 |X\rangle|k\rangle|0\rangle\longmapsto
 |X\rangle|k+f_{\mathrm{int}}(X)\bmod2^{33}\rangle|0\rangle.
\]
The trusted base is the mirrored gate semantics, coefficient data, canonical/bitsliced evaluators, Python/NumPy and asserted segmentation; this is a computational certificate, not a proof-assistant formalization.

\begin{table}[htbp]\centering\small
\caption{HHL verification inventory: twelve exact logical lemmas and a separate numerical accuracy check. Enumeration exhausts the stated domain; cancellation holds for all operands.}\label{supp:tab:hhlproof}
\begin{tabularx}{\linewidth}{@{}>{\raggedright\arraybackslash}Xrlr@{}}\toprule
Statement & Checks & Method & Points\\\midrule
\multicolumn{4}{@{}l}{\emph{Exact logical lemmas (12)}}\\

$X$ unchanged; unconditioned phase zero; all 902 frees clean; 845 of 899 phase coefficients cancel & 1 & Syntactic & ---\\
The other 54 phase coefficients vanish (support $X_9,\dots,X_{20}$) & 1 & Enumeration & 4,096\\
Tape equals the secure tape; all 1,163 writes to $k$ lie in the eight adder ranges & 1 & Syntactic & ---\\
Constant-term adder and seven Booth rows: $k\mathrel{+}=\pm(a+c_{\mathrm{in}})$, frame restored & 8 & Cancellation & ---\\
Addends depend only on $X$; their sum equals the integer model $f_{\mathrm{int}}(X)$ & 1 & Enumeration & 2,031,616\\
\addlinespace\multicolumn{4}{@{}l}{\emph{Numerical accuracy check (1; not a logical proof)}}\\
Integer increment versus 45-digit target; no certified reference enclosure & 1 & Numerical & 2,031,616\\
\bottomrule\end{tabularx}
\end{table}

The separate exhaustive 45-digit comparison finds a maximum error of approximately $417.45$ accumulator units ($9.719475\times10^{-8}$\,rad) at $X=151614$, below tolerance $429.4967296$ units. No rigorous transcendental-reference error bound or certified interval enclosure is available. A separate Qualtran kickmix conversion~\citep{khattar2026} reproduces counts and bit-for-bit outputs on $10^6$ random inputs, with zero sampled input, phase or workspace errors. Three proof faults and five simulator fault classes were detected; sampled checks remain distinct from the logical proof.

\paragraph{Signed-Booth circuits.} The same lemma structure, adapted to the combined first Booth rows, proves the logical action of the $987\times113$ and $972\times115$ signed-Booth circuits, with twelve lemmas each, exhaustive simulator agreement on all inputs and all injected faults detected. Their exhaustive 45-digit comparison gives a maximum error of 425.38 units ($9.90\times10^{-8}$\,rad) at $X=2{,}016{,}469$, a margin of 4.12 units against 12.05 for the Booth-polynomial circuit. The other three signed-Booth circuits pass the exhaustive output check without proofs.

Corollary~\ref{cor:inverse} supplies the logical inverse on clean workspace, retaining Toffoli count and peak qubits and adding 66 $X$ gates. Appendix~\ref{app:inverse-check} gives the bounded implementation replay. Controlled use and complete HHL integration remain unresolved.
\FloatBarrier

\subsection{Bias diagnostic and digital disagreement}
\label{app:bias-diagnostic}
Substituting the lattice probability into Eq.~\eqref{eq:bias} gives a normalized-bias bound of about $1.00625\times10^{-5}$ for the factor-sum circuit, a numerical evaluation rather than a confidence statement; a zero-in-$10^6$ independent-reference study gives about $1.30\times10^{-5}$ at one-sided 95\% confidence. The lattice bias of the hardened circuit is about four times the factor-sum magnitude but remains well inside the tolerance.
The lattice bias allowance for omitted and call-disagreement mass is below $10^{-10}$ contract units; it excludes numerical roundoff and is not an enclosure of all numerical error. Sampled intervals use a normal approximation: missing rare wrong outputs can omit their variance and invalidate nominal coverage.

\begin{figure}[htbp]\centering\begin{tikzpicture}
\begin{axis}[paperaxis,width=0.72\linewidth,height=4.4cm,xmin=-2.80,xmax=-2.55,ymin=-0.5,ymax=2.5,
ytick={0,1,2},yticklabels={{Sample, $10^6$},{Monte Carlo, $5{\times}10^8$},{Lattice}},
xlabel={Signed price bias ($10^{-6}$ contract units)},xmajorgrids=true,xtick distance=0.05,
xticklabel style={/pgf/number format/fixed,/pgf/number format/precision=2}]
\addplot[Sea!60,dashed] coordinates {(-2.625376361,-0.5) (-2.625376361,2.5)};
\draw[Ink,thick,{|[width=5pt]}-{|[width=5pt]}] (axis cs:-2.725404291,0) -- (axis cs:-2.630715249,0);
\addplot[only marks,mark=o,mark size=2.6pt,Ink,thick,fill=white] coordinates {(-2.678059770,0)};
\draw[Ink,thick,{|[width=5pt]}-{|[width=5pt]}] (axis cs:-2.726722590,1) -- (axis cs:-2.598776495,1);
\addplot[only marks,mark=o,mark size=2.6pt,Ink,thick,fill=white] coordinates {(-2.662749542,1)};
\addplot[only marks,mark=*,mark size=3pt,Sea] coordinates {(-2.625376361,2)};
\node[font=\footnotesize,anchor=west] at (axis cs:-2.613376361,2) {$-2.6254$};
\end{axis}\end{tikzpicture}
\caption{Factor-sum signed price bias. Filled point/line: numerical lattice value with bounded omission, without a certified enclosure. Hollow points/bars: sample means and nominal 95\% normal-approximation intervals; the million-path interval misses the lattice value. Coverage is not established.}
\label{fig:bias}\end{figure}
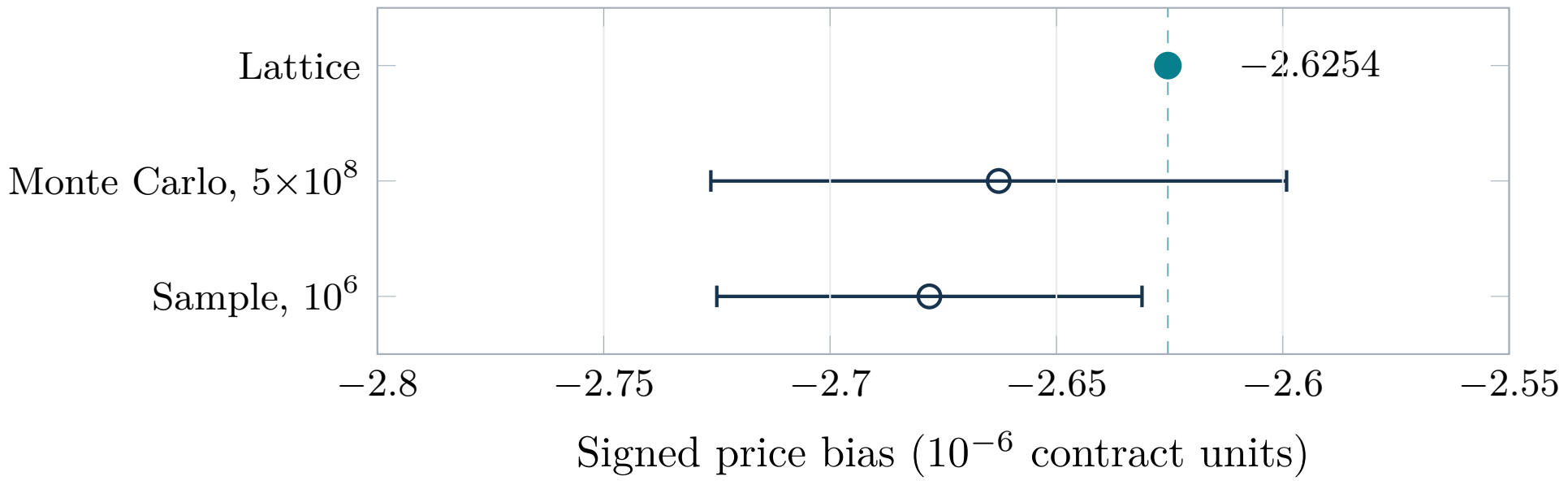

The million-path factor-sum sample contained no rare wrong outputs, while the separate $5\times10^8$-path study contained 25. Figure~\ref{fig:bias} shows the resulting interval discrepancy; agreement with the lattice does not certify coverage. Rare-event uncertainty in the log-return reference's sampled bias also remains unresolved. Exact figure data and full stored values remain in the numerical artifact.

\paragraph{Digital disagreement and angle error.}
The basis-action disagreement probability in the $2\sqrt{\varepsilon}$ trace-distance bound of \citet{khattar2026} differs from $p_{\mathrm{fail}}$: even a one-unit digital-word error gives an orthogonal basis word. The effective-payoff inequality in Section~\ref{sec:price} instead uses angle tolerance and exceptional-set probability, with numerical substitution subject to the lattice qualifications.

The manifest identifies a separate gate-level replay of both circuits on the same $10^6$ independent-reference paths, with zero incoming accumulators and fresh measurement seeds; no increment differences or sampled input, phase or workspace errors were observed.

\FloatBarrier
\label{budget:B-proofs-validation:end}
\section{TARF exceptional-event analysis}
\label{app:tarf-exceptions}
\subsection{Selected boundary inputs}
\label{sec:tarfboundary}
A protocol fixed before fixture construction selected 5,218 paths using an independent 60/90-digit reference from the contract constants: all 1,024 first-two-index pairs, reachable knock-out/strike neighbours, window edges, near-cap, deep and tail paths, plus 1,000 loading-law sanity paths. Both circuits used secure build/recounts and the unchanged checker with two measurement histories giving bit-for-bit identical increments; reference and task verdicts agree.

Inside the windows, no tested knock-out or strike decision is wrong. Every selected pre-termination lower crossing gives a wrong output; angle-stage range limits occur only after such crossings. Near-cap failures (Table~\ref{tab:tarfboundary}) produce angle errors of $0.010$--$0.52$\,rad, sometimes amplified by later negative payments. Diagnosis uses reference flags and counterfactuals, without a gate-level trace. These selected inputs are not additional holdout draws. A separate million-path loading-law sample reaches cap margin $7.0\times10^{-6}$ without establishing a circuit-failure rate there.

\begin{table}[htbp]\centering\small\setlength{\tabcolsep}{4pt}
\caption{Selected-input TARF outcomes. \emph{Window cut}: below the circuit's index window before termination. \emph{Cap tie}: accrual within payment resolution of the cap. Largest error is over paths within the $10^{-5}$\,rad tolerance. Counts are diagnostic, not failure rates.}\label{tab:tarfboundary}
\begin{tabular}{lrrrrrr}\toprule
Circuit & Violations & \begin{tabular}[b]{@{}r@{}}Window\\cut\end{tabular} & \begin{tabular}[b]{@{}r@{}}Cap\\tie\end{tabular} & Other & \begin{tabular}[b]{@{}r@{}}Phase/ancilla/\\input errors\end{tabular} & \begin{tabular}[b]{@{}r@{}}Largest\\error (rad)\end{tabular}\\\midrule
Windowed circuit & 1,038 & 991 & 47$^a$ & 0 & 0/0/0 & $6.45\times10^{-6}$\\
Lookup reference & 627 & 625 & 2 & 0 & 0/0/0 & $3.93\times10^{-6}$\\
\bottomrule\end{tabular}
\par\smallskip{\footnotesize $^a$ Includes 8 cases whose mechanism was identified only after the replay.}
\end{table}

\subsection{Independent TARF lower-window crossing bound}
\label{app:tarf-window}
The windowed circuit's lower thresholds for dates 1--26 are
\begin{align*}
(L_1,\ldots,L_{13})&=(0,6,16,27,38,50,63,75,88,101,114,127,140),\\
(L_{14},\ldots,L_{26})&=(153,167,180,194,208,221,235,249,263,276,290,304,317).
\end{align*}
Let $q_0(k)=\mathbf1_{k=0}$. With the index probabilities specified below, propagate
\begin{equation}
 r_t(k)=\sum_{j=0}^{31}p_jq_{t-1}(k-j),\qquad
 h_t=\sum_{k<L_t}r_t(k),\qquad
 q_t(k)=\mathbf1_{k\ge L_t}r_t(k).
 \label{eq:tarf-window-dp}
\end{equation}
Killing crossed mass makes the events counted by $h_t$ disjoint, so $\sum_{t=1}^{26}h_t$ is the unrestricted process's first-passage probability. Ignoring cap and knock-out termination gives a superset of lower crossings while the circuit is alive; this event bound does not cover other failure mechanisms.

Normalize the saved binary64 entries $\bar p_j$ exactly: $S=\sum_j\bar p_j$, $p_j=\bar p_j/S$. Common-denominator integer weights make convolution, killed mass and mass-conservation checks exact, with powers of the weight sum as denominators. The first-passage probability is $4.95104459856437\ldots\times10^{-7}$, agreeing with a separate float64 convolution. The raw 26-index product measure differs from the normalized crossing mass by at most $|S^{26}-1|<2.624\times10^{-15}$. Adding this allowance and rounding upward gives
\begin{equation}
 \Pr\{\exists t:K_t<L_t\}<\TarfWindowUpper.
 \label{eq:tarf-window-bound}
\end{equation}
The bound is rigorous for the explicitly normalized stored law, with the stated normalization allowance; it does not certify transcendental generation of the weights or a finite random-number generator. The artifact manifest identifies the exact arithmetic records.

\FloatBarrier

\subsection{Cap-tie probability analysis}
\label{app:tarf-captie}
Under a protocol fixed before sampling, this reference-only study draws $\TieN$ loading-law paths without running a circuit. At an alive date let $m_t$ be the pre-date accrual plus that date's payment minus the cap. Define $A_b=\{\exists\text{ alive }t:|m_t|<b\}$. A path is in $E_b\subseteq A_b$ if either flipping one such cap decision or persistently shifting the cap threshold just past that margin changes the 60-digit reference angle by more than $10^{-5}$\,rad. Let $L_b\subseteq E_b$ require a later price below 15 before reaching 20.

\begin{table}[htbp]\centering\small
\caption{Reference-defined near-cap events: approximate one-sided 95\% Clopper--Pearson upper endpoints from $\TieN$ paths, with every near-tie path recomputed at 60 digits. These concern $P(E_b)$, not circuit failure rates.}\label{tab:cap-events}
\begin{tabularx}{\linewidth}{@{}lr>{\raggedright\arraybackslash}X@{}}\toprule
Band $b$ & Upper endpoint & Relation to selected circuit fixtures\\\midrule
$10^{-4}$ & $\TieWin$ & Windowed wrong outputs at $|m_t|\le4.15\times10^{-5}$, none at $|m_t|\ge10^{-4}$\\
$5\times10^{-5}$ & $\TieWinHalf$ & Narrower comparison band for the windowed circuit\\
$2\times10^{-7}$ & $\TieRef$ & Lookup reference wrong at $3.3\times10^{-8}$, correct on all consequential ties with $|m_t|\ge1.7\times10^{-7}$\\\bottomrule
\end{tabularx}
\end{table}
At $b=10^{-4}$, $\TieCons$ of $\TieNear$ near-tie paths reach $E_b$: $\TieConsExit$ reach $L_b$, each with maximum flip-or-shift angle change at least $\TieExitMinDtheta$\,rad; the other $\TieConsNoExit$ remain in $[15,20)$, with changes up to approximately $\TieNoExitMaxDtheta$\,rad but above tolerance. A second method sums the last index and future after each tie exactly, enumerating or sampling earlier prefixes. It agrees within three combined standard errors for $P(A_b)$ and $P(L_b)$ in every tested band, but does not independently compute $P(E_b)$. Repeated lattice margins make $P(E_b)$ nonlinear in $b$; bands are not extrapolated.

A cap-error bound requires the unproved containment that every circuit cap-tie error implies $E_b$. Boundary inputs suggest this relationship but do not prove it from circuit rounding. The windowed circuit rounds payments to $2^{-15}$ ($2^{-14}$ from date 14); 26 roundings could accumulate to about $10^{-3}$. At $b=10^{-3}$ ($\TieSensN$ paths), the upper endpoint is approximately \TieSens{}, above the $10^{-5}$ task budget. Thus a rounding argument is essential; more boundary tests cannot establish the containment. Conversely, 35 selected consequential paths leave the windowed circuit correct, so membership in $E_b$ does not imply circuit failure.

Under that containment at the table's $10^{-4}$ windowed and $2\times10^{-7}$ lookup bands, combining the coarse $5\times10^{-7}$ lower-window bound with the cap-event endpoint gives at most \TieWinSum{} for those two windowed-circuit mechanisms. Using the lookup reference's source-stated $10^{-10}$ cut per date and side, not re-estimated here, gives \TieRefSum{}. These are below the task budget and million-path holdout bounds, but remain conditional mechanism bounds: other failures are not excluded, and no unconditional total circuit-failure bound follows.

\FloatBarrier
\label{budget:C-tarf-events:end}

\section{Search provenance and reproducibility}
\label{app:search-protocols}
The project was underway when \citet{gu} appeared; we then adopted its evaluator-guided organization. Table~\ref{tab:provenance} summarizes the retained design runs.

\begin{table}[!htb]\centering\footnotesize
\caption{Search budgets, access and selection. These runs do not establish repeated-search reliability.}\label{tab:provenance}
\setlength{\tabcolsep}{4pt}
\begin{tabularx}{\linewidth}{@{}>{\raggedright\arraybackslash}p{1.8cm}*{3}{>{\raggedright\arraybackslash}X}@{}}\toprule
Study & Budget / trials & Access / assistance & Selection / qualification\\\midrule
Autocallable / HHL & 4 / 5 programs; no predeclared stop & Full repository; first program in ideation; HHL handwritten hints & Lowest retained secure product; assisted runs\\\addlinespace
TARF & 7 sessions: ideation, 4 designs, 2 rewrites; 6 programs & Sanitized task/harness; no candidates/history; index-sum model read by 4 sessions; no extra hints & Lowest secure product; preregistered million-path holdout; programs share one design lineage\\\addlinespace
Second setting & Ideation + 3 designs (90\,min/100 turns); 2 circuits, 1 empty trial & Isolated sessions; task/model/primitives and helper study accessible; no extra hints & Protocol eligibility/seeds/sizes; failed/unfilled slots retained in joint confidence family\\\addlinespace
References / controls & Interactive references; 6 control configurations; 3 fixed transfers & References pre-search; TARF reference excluded; AI-assisted control generator, no model queries in fitting & Valid secure-screening minimum product; ties by qubits then ID; transfers fixed before scoring\\\addlinespace
Optimiz\-ation stage & Coordinated agents; no session budget & Full repository, including retained circuits, proofs and selected stress inputs; offline fitting on sampled paths & Lowest secure product and accuracy-hardened alternatives; one recorded holdout seed\\\bottomrule
\end{tabularx}
\end{table}
\FloatBarrier

\paragraph{Material deviations.}
\label{app:second-search}
Early searches were assisted; factor-sum reused an earlier put stage. TARF ideation exposed to a prior-design protocol was voided, retained and rerun in isolation. Its programs are refinements, not independent replications; two sessions abandoned their family ideas. Three sessions read their saved output outside the workspace under an audit allowance added during the run. TARF and second-setting searches had no additional run-specific design hints, with task files accessible. Second-setting audits found no repository/prior-result/other-session reads; tool exceptions are recorded. The validation driver was written after search scores were visible, then frozen before protocol-defined final tests. Early exact model identifiers and verbatim HHL hints are missing; later sessions record the model identifier and settings.

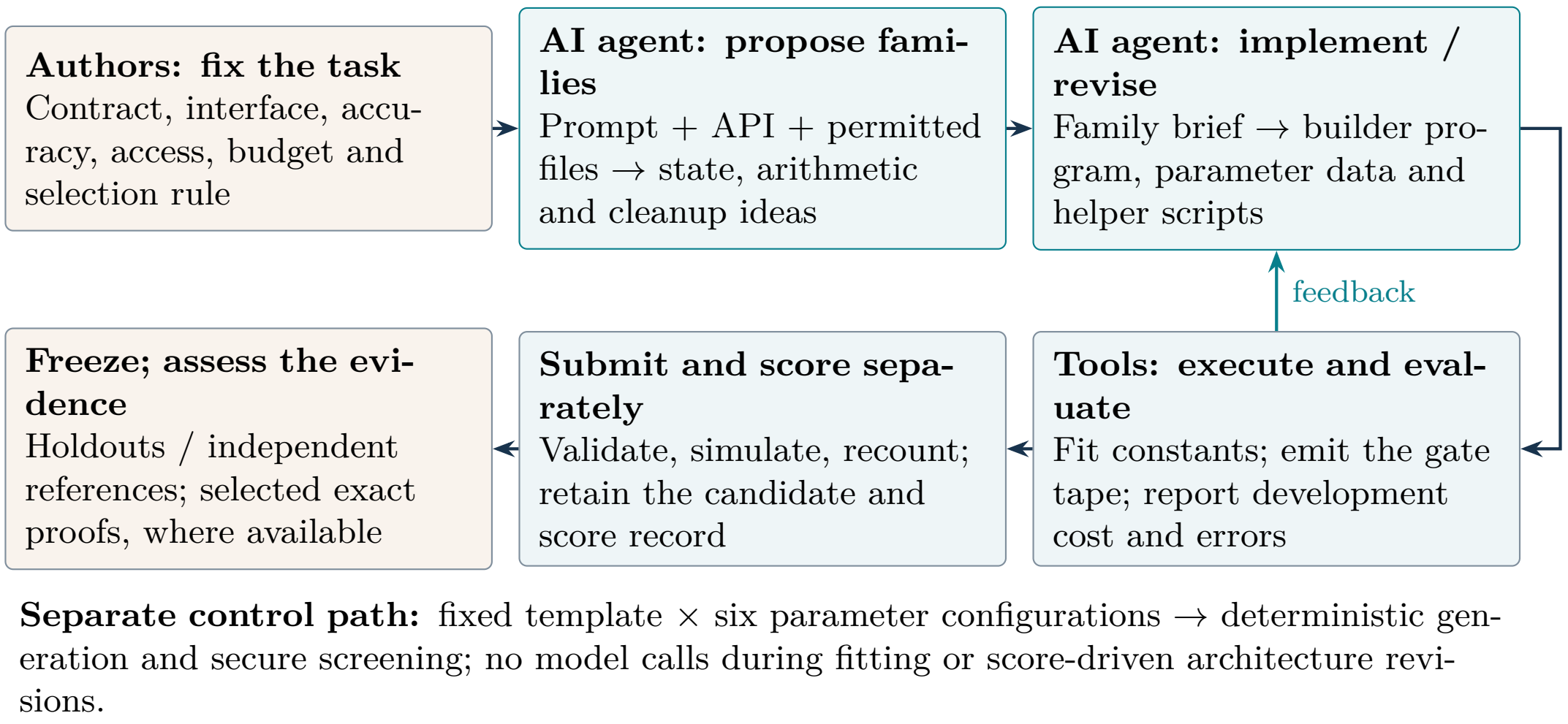
\begin{figure}[!htbp]
\centering
\begin{tikzpicture}[x=1cm,y=1cm,
 wf/.style={draw=Ink!55,rounded corners=2pt,align=left,text width=4.05cm,minimum height=1.85cm,inner sep=5pt,font=\footnotesize},
 model/.style={wf,draw=Sea,fill=Sea!7},
 tool/.style={wf,fill=Pale},
 wfarr/.style={-{Stealth[length=2mm]},draw=Ink,thick}]
\node[wf,fill=Gold!8] (task) at (0,0) {\textbf{Authors: fix the task}\\Contract, interface, accuracy, access, budget and selection rule};
\node[model] (idea) at (4.65,0) {\textbf{AI agent: propose families}\\Prompt + API + permitted files $\to$ state, arithmetic and cleanup ideas};
\node[model] (code) at (9.3,0) {\textbf{AI agent: implement / revise}\\Family brief $\to$ builder program, parameter data and helper scripts};
\node[tool] (eval) at (9.3,-2.9) {\textbf{Tools: execute and evaluate}\\Fit constants; emit the gate tape; report development cost and errors};
\node[tool] (score) at (4.65,-2.9) {\textbf{Submit and score separately}\\Validate, simulate, recount; retain the candidate and score record};
\node[wf,fill=Gold!8] (study) at (0,-2.9) {\textbf{Freeze; assess the evidence}\\Holdouts / independent references; selected exact proofs, where available};
\draw[wfarr] (task)--(idea);\draw[wfarr] (idea)--(code);
\draw[wfarr] (code.east)--++(.36,0)|- (eval.east);
\draw[wfarr,draw=Sea] (eval.north)--node[right,font=\scriptsize,text=Sea]{feedback} (code.south);
\draw[wfarr] (eval)--(score);\draw[wfarr] (score)--(study);
\node[font=\footnotesize,align=left,text width=13.45cm,anchor=north] at (4.65,-4.15) {\textbf{Separate control path:} fixed template $\times$ six parameter configurations $\to$ deterministic generation and secure screening; no model calls during fitting or score-driven architecture revisions.};
\end{tikzpicture}
\caption{Roles and information flow in our evaluator-guided design process, following the organization of \citet{gu}. The feedback loop is within development; final evidence is separate. Access, assistance and reuse vary by run (Appendix~\ref{app:search-protocols}); family labels do not imply independent implementations. Appendix~\ref{app:search-protocols} traces the second-setting lookup circuit through this workflow.}
\label{fig:design-workflow}
\end{figure}

\paragraph{Illustrative design session.} One retained example is the second-setting lookup lattice-form circuit. Ideation returned three families; one combined fitted per-asset integer forms with per-factor lookups. Starting from a blank candidate, the agent wrote the circuit, a lattice fitter and an emulator. Development failures exposed a fitter error: a threshold had been allowed to split equal integer-form values. The agent corrected the tie scoring and refitted asset 2, replacing an own-factor weight of 1993 by 11932; the final form is $4616K_{t,0}+3203K_{t,1}+11932K_{t,2}$. The submitted circuit uses two 32-entry lookups and factor-2 shift-adds and scores 13,661 Toffoli gates and 524 qubits. The audit records no human repairs in this session.

\paragraph{Optimization stage.} The barrier-centred, narrow-window and signed-Booth circuits came from a later stage without a session budget or sanitized workspace. Its coordinated agents (task-specific designers, a read-only verifier and a stall monitor) could read the full repository, including the retained circuits, proofs, validation records and selected stress inputs, and fitted constants offline on $1.6\times10^7$ to $10^8$ sampled loading-law paths. These circuits therefore build on the search results and do not measure the reliability of the search protocol.

\FloatBarrier
\label{budget:D-provenance:end}

\end{document}